\documentclass[iop,twocolappendix]{emulateapj}
\usepackage{epsfig}
\usepackage{mathptmx} 
\usepackage{graphicx}
\usepackage[breaklinks=true]{hyperref} 

\shorttitle{Dwarf Galaxies with Radiation Feedback. I}
\shortauthors{J. Kim et al.}

\begin{document}

\title{Dwarf Galaxies with Ionizing Radiation Feedback. I: Escape of Ionizing Photons}

\author{Ji-hoon Kim \altaffilmark{1,2}}
\email{me@jihoonkim.org}
\author{Mark R. Krumholz \altaffilmark{1}}
\author{John H. Wise \altaffilmark{3}} 
\author{Matthew J. Turk \altaffilmark{4}} 
\author{Nathan J. Goldbaum \altaffilmark{1}}
\author{Tom Abel \altaffilmark{2}}

\altaffiltext{1}{Department of Astronomy and Astrophysics, University of California, Santa Cruz, CA, USA}
\altaffiltext{2}{Kavli Institute for Particle Astrophysics and Cosmology, Stanford University, Stanford, CA, USA}
\altaffiltext{3}{Center for Relativistic Astrophysics, School of Physics, Georgia Institute of Technology, Atlanta, GA, USA}
\altaffiltext{4}{Department of Astronomy and Astrophysics, Columbia University, New York, NY, USA}

\begin{abstract}
We describe a new method for simulating ionizing radiation and supernova feedback in the analogues of low-redshift galactic disks. 
In this method, which we call star-forming molecular cloud (SFMC) particles, we use a ray-tracing technique to solve the radiative transfer equation for ultraviolet photons emitted by thousands of distinct particles on the fly.
Joined with high numerical resolution of 3.8 pc, the realistic description of stellar feedback helps to self-regulate star formation.  
This new feedback scheme also enables us to study the escape of ionizing photons from star-forming clumps and from a galaxy, and to examine the evolving environment of star-forming gas clumps.  
By simulating a galactic disk in a halo of $2.3 \times 10^{11} M_{\odot}$, we find that the average escape fraction from all radiating sources on the spiral arms (excluding the central 2.5 kpc) fluctuates between 0.08\% and 5.9\% during a $\sim$ 20 Myr period with a mean value of 1.1\%.
The flux of escaped photons from these sources is not strongly beamed, but manifests a large opening angle of more than $60^{\circ}$ from the galactic pole.  
Further, we investigate the escape fraction per SFMC particle, $f_{\rm esc}(i)$, and how it evolves as the particle ages.  
We discover that the average escape fraction $f_{\rm esc}$ is dominated by a small number of SFMC particles with high $f_{\rm esc}(i)$.  
On average, the escape fraction from a SFMC particle rises from 0.27\% at its birth to 2.1\% at the end of a particle lifetime, 6 Myrs.  
This is because SFMC particles drift away from the dense gas clumps in which they were born, and because the gas around the star-forming clumps is dispersed by ionizing radiation and supernova feedback.  
The framework established in this study brings deeper insight into the physics of photon escape fraction from an individual star-forming clump, and from a galactic disk.
\end{abstract}


\section{Introduction}\label{sec-III:1}

Lying at the center of a web of processes that drive cosmic evolution is galactic star formation.  
Decades of observations have yielded a number of empirical rules about how it operates. 
The accumulated observational knowledge considerably exceeds our physical understanding, and as a result simulations and semi-analytic models of galaxy formation, reionization, and interstellar medium (ISM) dynamics often rely on {\it subgrid} models of uncertain validity for converting gaseous material into stars and then injecting energy, momentum, and newly-synthesized elements back into the ISM.

Theoretical understanding of star formation has been hampered primarily by the fact that star-forming gas clumps do not exist in isolation.  
Rather, they are a part of a larger {\it galactic} structure. 
Molecular clouds are strongly associated with peaks of the atomic hydrogen distribution in galaxies \citep[e.g.][]{2010MNRAS.406.2065H}, and both direct kinematic measurements and indirect statistical evidence suggest that giant molecular clouds (GMCs) spend much of their lifetime accreting mass from their surrounding atomic envelopes \citep[e.g.][]{2009ApJ...705..144F}. 
Gas accretion alters molecular clouds by providing mass to replace that ejected by stellar feedback, and by adding energy to offset the decay of turbulence and stall gravitational collapse \citep[e.g.][]{2010ApJ...715.1302V, 2011ApJ...738..101G}. 
Consequently, the internal processes that drive and regulate star formation in molecular clouds cannot be understood outside the larger galactic context. 

Needless to say, numerical investigations of star-forming gas clumps in a galactic context pose a problem of inherently multi-scale nature. 
In local galaxies, GMCs have typical surface densities of $\Sigma \sim 100 \,\, M_{\odot} {\rm pc}^{-2}$ \citep{2008ApJ...686..948B}, and star formation occurs inside even denser clumps ($\Sigma \sim 1000  \,\, M_{\odot} {\rm pc}^{-2}$; \citealt{1997ApJ...476..730P, 2010A&A...519L...7L}). 
In contrast, GMCs' atomic hydrogen envelopes are less dense ($\Sigma \sim 10  \,\, M_{\odot} {\rm pc}^{-2}$), and the surrounding ISM can be even more diffuse ($\Sigma \lesssim 1  \,\, M_{\odot} {\rm pc}^{-2}$).
Indeed, the large dynamic range of the problem has prohibited theorists from numerically resolving individual star-forming events in a simulation of a long-term galactic evolution. 

\begin{figure*}[t]
\epsscale{1.13}
\plotone{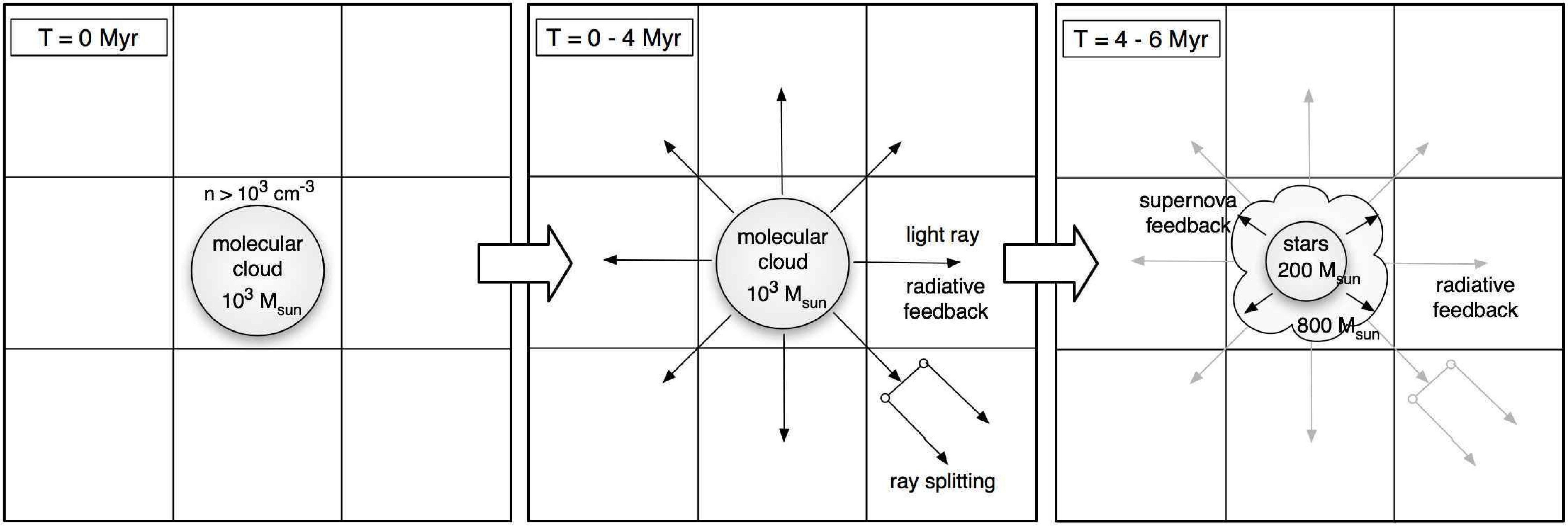}
    \caption{Two dimensional schematic overview of the life cycle of star-forming molecular cloud (SFMC) particle formation and two different channels of feedback.  
{\it Left:} formation of SFMC particles described in \S \ref{sec-III:2-formation}.  The numerical resolution of 3.8 pc is in accord with the Jeans length for a gas clump of $n=1000$ ${\rm cm^{-3}}$ at $\sim$100 K, at which point a molecular cloud particle of $1000 \,\,M_{\odot}$ spawns. 
{\it Middle:} radiation feedback described in \S \ref{sec-III:2-RF}.  Rays of ionizing photons carrying the monochromatic energy of 16.0 eV are adaptively traced for particles of age $T=0$ to 6 Myr.  
{\it Right:} supernova feedback described in \S \ref{sec-III:2-TF}.  Along with the radiation, the thermal energy by supernovae explosion is injected into the gas cell in which a SFMC particle of age $T=4$ to 6 Myr resides.
\label{fig:RTF}}
\end{figure*}

A variety of methods have been proposed to limit the dynamic range of a galactic ISM simulation.  
Simply speaking, these are attempts to model star formation and feedback at the numerical resolution one can afford.  
A popular choice among these is to use a subgrid model to estimate the pressurization of gas on large scales  ($\sim$ 100 pc) and impose a pressure floor at the calculated level \citep[e.g.][]{2005MNRAS.361..776S}. 
One may also impose artificial pressurization to stabilize the ISM at smaller scales \citep[e.g.][]{2010ApJ...720L.149T, 2010MNRAS.409.1088B}. 
While these pressure floor models successfully halt gravitational collapse and produce equilibrium galaxies, the treatment of subgrid physics is model-dependent and must be tuned to match observations. 
Another option is to allow the gas to collapse unimpeded and then assume that the gas that exceeds the highest resolvable density forms a bound structure \citep[Lagrangian sink particle; e.g.][]{2005ApJ...626..823L}. 
However, this approach may overestimate the amount of gas in the dense ISM, especially in simulations that do not include proper feedback. 
A related approach is to allow unimpeded collapse but not to put all the collapsing gas into a sink particle, and instead rely on star formation recipes \citep{2008ApJ...684..978S, 2009ApJ...700..358T, 2011MNRAS.413.2935D}.
This approach, however, necessarily involves a violation of the Jeans resolution condition at the finest level, with unknown consequences \citep{1997ApJ...489L.179T}.
Hence, one might arguably claim that there has been no numerical simulation that can self-consistently couple the properties of a galaxy to the internal properties of star-forming gas clumps.  
At the same time, no simulation has formed a population of numerically converged molecular clouds that agrees with the GMC populations observed in external galaxies.  
Because of these outstanding challenges, it is unclear how the properties of molecular clouds are imprinted with the properties of their host galaxies.

The goal of this work is to improve this situation by performing first-principles simulations of the behavior of star-forming gas clumps in a galactic context. 
In particular, we aim to address the physics of star-forming clumps in a galaxy formation simulation with ever-improving computational resolution.
As the numerical resolution of galaxy simulations dramatically improves, spatial resolution of $\sim$ pc is now frequently employed and tested by a number of authors \citep[e.g.][]{2010ApJ...720L.149T, 2010MNRAS.409.1088B, 2011MNRAS.417..950H}.  
Yet, the formation and feedback of star clusters in galaxy formation simulations have been described only {\it phenomenologically} at best. 
Such examples include placing a star cluster particle using an observed star formation density relation \citep[e.g. $\rho_{\rm SFR} \sim \rho_{\rm gas}^{1.5}$;][]{1959ApJ...129..243S, 1998ApJ...498..541K}, or stopping the cooling near a star cluster particle right after its birth in an attempt to mimic the supernova energy injection \citep[e.g.][and references therein]{2012MNRAS.422.1231G}.  
These phenomenological prescriptions may have worked effectively and efficiently at low resolution.  
However, a careful investigation is greatly needed to depict the realistic {\it physics} of star-forming gas clumps in a simulation that resolves the molecular cloud scale of $\Delta\, x \sim {\rm pc}$ with a corresponding timestep of $\Delta\, t \sim 3 \times 10^4 \,\, {\rm yr}$ at $10^4 \,\,{\rm K}$.
One should note that this fine timestep now starts to resolve the lifetime of typical molecular clouds and massive stars, placing many high-resolution simulations in great need of a {\it temporally-resolved} stellar feedback model \citep[for recent attempts to tackle this challenge, see][]{2011MNRAS.417..950H, 2012arXiv1208.0002S}.  

In light of these developments, we take advantage of the high dynamic range made possible by an adaptive mesh refinement (AMR) technique to simulate the behavior of $\lesssim$10000 star-forming molecular cloud (SFMC) particles {\it and} their embedding galaxy in a single self-consistent numerical framework. 
We have implemented a sophisticated description of stellar feedback by combining the ultraviolet radiation from star clusters and the thermal energy injection by supernova explosion.   
The numerical resolution of 3.8 pc helps to resolve the star-forming environment on a low-redshift analogue of a galactic disk in the halo of $2.3 \times 10^{11} M_{\odot}$, and to describe the capture of stellar radiation energy.  
Our work fundamentally differs from the representation of stellar feedback in \cite{2011MNRAS.417..950H} and their subsequent studies.  
Specifically, we do not artificially pressurize the ISM at small scales, nor rely on a subgrid model to handle the stellar radiation feedback. 
Instead, we actually solve the transfer equation on the fly. 
This means that our approach does not rely on free parameters to describe the interaction of radiation and the gas, and also that we can address questions that previous studies cannot. 
In this paper we focus on one of these: by solving the transfer equation, we are able to study the escape of ionizing radiation from galaxies self-consistently, with the ionizing radiation actually included in the simulation rather than added in post-process. 
Future work will focus on other aspects of our simulations, including regulation of star formation by ionizing radiation feedback, and mock observations of ionized gas tracers such as H$\alpha$ emission.

In a series of papers, we will discuss the state-of-the-art model that describes feedback from SFMC particles and how it enhances our understanding of galactic star formation by significantly expanding the usage of numerical simulations.   
The first paper of the series is organized as follows.  
The physics in the simulation code is described in \S \ref{sec-III:2}, and the initial condition of the numerical experiments is explained in \S \ref{sec-III:3}.  
\S \ref{sec-III:4} overviews the performed simulations of dwarf-sized galactic disks with radiation feedback, and \S \ref{sec-III:5} is devoted to the results of one of our experiments, with a strong emphasis on the escape of ionizing photons.  
Particular attention is paid to the evolution of the escape fraction per SFMC particle. 
Assembled in \S \ref{sec-III:6} are the summary and conclusions.  

\section{Physics In The Code}\label{sec-III:2}

We utilize the Eulerian adaptive mesh refinement  {\it Enzo}, a parallel hybrid N-body radiation hydrodynamics code  \citep{1997ASPC..123..363B, 1999ASSL..240...19N, 2007arXiv0705.1556N, 2011MNRAS.414.3458W}. 
The code contains several basic modules. 
The hydrodynamics module solves the Euler equations of compressible gas dynamics using either a piecewise-parabolic method or a ZEUS scheme. 
The gravity module computes the gravitational potential generated by the fluid, collisionless dark matter and stellar particles.  
The latter two are discretized onto the grid using an adaptive particle-mesh method, allowing {\it Enzo} to solve the Poisson's equation for the potential via fast Fourier transform and/or multigrid solvers. 
The radiation transfer module propagates ionizing photons through a spatially-adaptive tree of rays drawn around each source, and updates the thermal and chemical states of the gas as a result of its interaction with these photons. 
All these modules operate on top of an AMR hierarchy that allows users to specify arbitrary refinement conditions.
In addition to these basic modules, our version of {\it Enzo-2.1}\,\footnote{http://enzo-project.org/} contains all relevant physics previously considered in galactic simulations \citep[e.g.][]{2009ApJ...694L.123K, 2011ApJ...738...54K} as well as other enhanced  physics discussed in detail below.
As a result of the wide range of physics included, we are well-poised to investigate the physics of star-forming clumps in a galactic context.  

\subsection{Hydrodynamics, Refinement, and Chemistry} \label{sec-III:2-hyd}

We use the ZEUS hydrodynamics module included in {\it Enzo} to solve the Euler equations for the collisional fluid \citep{1992ApJS...80..753S, 1992ApJS...80..791S, 1994ApJ...429..434A}.
Dark matter and stars are treated as collisionless particles.  
The grids are adaptively refined by factors of 2 in each axis on gas and particle overdensities. 
The mass thresholds, $M_{\rm ref}$, above which a cell is refined are functions of a refinement level {\it l} as
\begin{eqnarray}
M_{\rm ref, gas}^{l} &=& 2^{-0.820 l} \,M_{\rm ref, gas}^0 \,\,, \\
M_{\rm ref, part}^{l} &=& 2^{-0.533 l} \,M_{\rm ref, part}^0 
\end{eqnarray}
for gas and particles, respectively, with $M_{\rm ref, gas}^0 = M_{\rm ref, part}^0 = 4.87\times 10^6 \,M_{\odot}$ for a $32^3$ root grid.  
This makes the simulation super-Lagrangian, refining grids more on small scales.
At the finest level $l=13$ of 3.8 pc resolution, a cell would have been further refined if more than $3000 \,\,M_{\odot} $ in gas or $40000 \,\,M_{\odot} \simeq 2.5 \,\, M_{\rm DM}$ in particles.  
This numerical resolution is in accord with the Jeans length for a dense gas clump of $n=1000$ ${\rm cm^{-3}}$ at $\sim$ 100 K, at which point a corresponding Jeans mass of $2000 \,\,M_{\odot}$ collapses to spawn a SFMC particle (see \S \ref{sec-III:2-formation}).  
Because 1000 to 1500 $M_{\odot}$ is instantly removed from the cell every time a particle is created, the gas mass in the finest cell never reaches the refinement threshold.

We choose the hydrodynamic Courant-Friedrichs-Lewy safety number of 0.3 so that the marching timestep is less than the time for the advecting wave to reach the adjacent cell.  
We make use of {\it Enzo}'s multi-species non-equilibrium chemistry module to track the reactions among six species (H, ${\rm H}^+$, He, ${\rm He}^+$, ${\rm He}^{++}$, ${\rm e}^-$), and also employ its cooling module to calculate radiative losses due to collisional excitation cooling, collisional ionization cooling, recombination cooling, Bremsstrahlung cooling, and CMB Compton cooling for hydrogen and helium \citep{1997NewA....2..209A}. 
Added to the primordial cooling rate is the metallicity-dependent metal cooling rate $\Delta \Lambda(Z) = \Lambda_{\rm net}(Z) - \Lambda_{\rm net}(0)$ in gas above $10^4$ K, where $\Lambda_{\rm net}$ is the net cooling rate tabulated in \cite{1993ApJS...88..253S}.  
Furthermore, cooling by metals below $10^4$ K uses the cooling rate approximation of \cite{2002ApJ...564L..97K} (see however \cite{2010ApJ...715.1302V} for an important typo).
Photoionization heating by the metagalactic ultraviolet background of quasars and galaxies is also taken into account without considering the effect of self-shielding \citep{1996ApJ...461...20H, 2001cghr.confE..64H}.  

\begin{figure}[t]
\epsscale{1.15}
\plotone{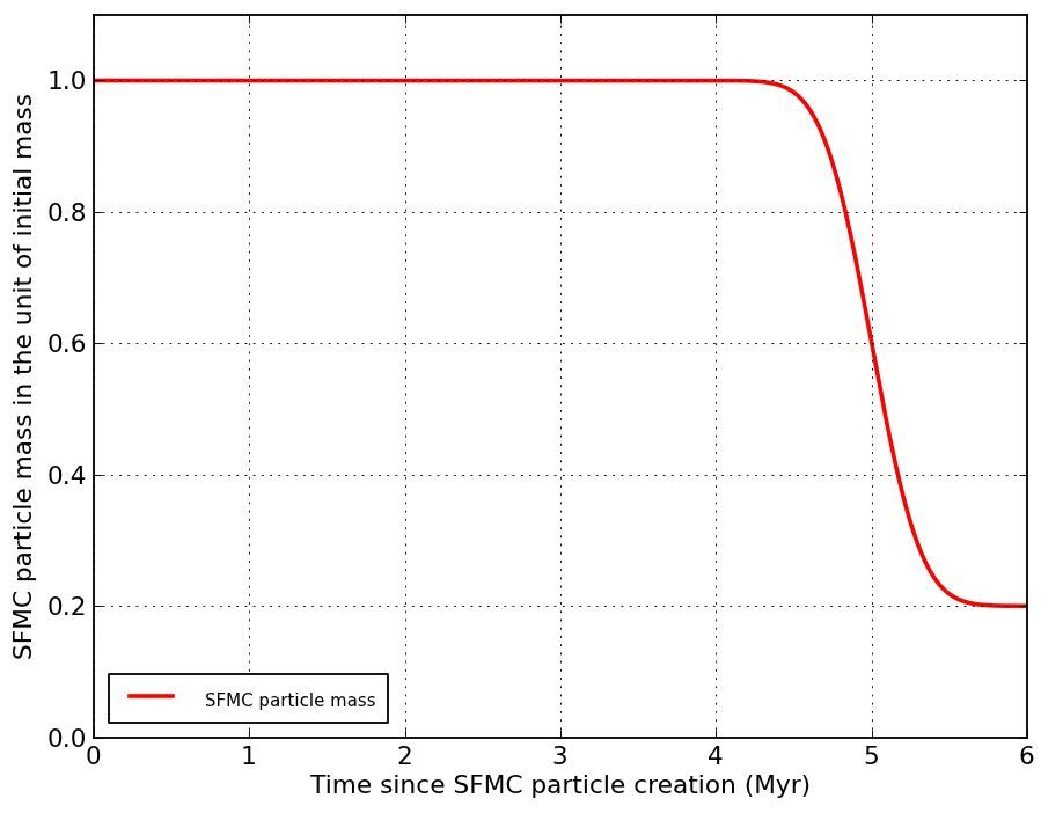}
    \caption{The evolution of SFMC particle mass, $M_{\rm MC}$, as a function of particle age $T$.  See \S\ref{sec-III:2-TF} and Eq.(\ref{eq:SFMC_mass}) for detailed explanations.
\label{fig:MC_mass_description}}
\end{figure}

\subsection{Formation of Star-forming Molecular Cloud (SFMC) Particles} \label{sec-III:2-formation}

We describe the model of formation and feedback of SFMC particles in the remainder of \S\ref{sec-III:2}.
Our SFMC particle formation is based on \cite{1992ApJ...399L.113C} with modifications similar to \cite{2011ApJ...738...54K}.
With an efficiency of $\epsilon_* = 0.5$, the finest cell of size $\Delta\, x = 3.8 \,\,{\rm pc}$ and gas density $\rho_{\rm gas}$  produces a SFMC particle of initial mass $M_{\rm MC}^{\rm init} = \epsilon_* \rho_{\rm gas} \Delta \,x^3$ when 
\begin{itemize}
\item[{\it (a)}] the proton number density exceeds the threshold $n_{\rm thres}=1000$ ${\rm cm^{-3}}$, 
\item[{\it (b)}] the velocity flow is converging, 
\item[{\it (c)}] the cooling time $t_{\rm cool}$ is shorter than the gas dynamical time $t_{\rm dyn}$ of the cell ($\sim 1 \,\, {\rm Myr}$), and 
\item[{\it (d)}] the particle produced has at least $M_{\rm MC}^{\rm init}  >  M_{\rm thres}  = 1000 \,\,M_{\odot}$.
\end{itemize}
As a result, the gas in the finest cell is instantly converted into a particle as soon as the cell has accumulated more than $M_{\rm thres} / \epsilon_*= 2000 \,\,M_{\odot}$, the Jeans mass at $n = 1000$ ${\rm cm^{-3}}$ at $\sim$ 100 K (see the left panel of Figure \ref{fig:RTF}).   
This guarantees that a SFMC particle is created before an unresolved gas clump interferes with the consistency of hydrodynamics (see \S \ref{sec-III:2-hyd}).  
Because we deposit a particle when a gas cell of a typical molecular cloud size actually becomes Jeans unstable, the particle in our simulation represents a {\it star-forming molecular cloud} that is self-gravitating, is thus decoupled from the gas on the grid.
Additionally, two SFMC particles merge in a way that conserves mass and momentum if they are separated by less than 5 pc in space, and $3 \times 10^4$ yr in age.  
This is to reduce the number of active SFMC particles and thereby expedite the radiative transfer calculation.

\begin{table*}[t]  
\caption[Simulation suite description]{Simulation Suite Description}
\centering     
\begin{tabular}{l  l  ||  c  c  c  c } 
\hline\hline   
\multicolumn{2}{c ||}{Physics\tablenotemark{a}} & MC-TF  & MC-RTF  \\ [1ex] 
\hline      
Star-forming molecular cloud (SFMC) particle formation       & (See \S \ref{sec-III:2-formation}) &\textcircled{}&\textcircled{}\\    
SFMC feedback: supernova explosion      & (See \S \ref{sec-III:2-TF}) &\textcircled{}&\textcircled{}\\    
SFMC feedback: ionizing radiation            & (See \S \ref{sec-III:2-RF}) &$\times$&\textcircled{}\\  [1ex] 
\hline
\end{tabular} 
\tablenotetext{1}{\scriptsize For detailed explanation, see the referenced section. $(\circ)$ included, $(\times)$ not included.}
\label{table-III:desc}  
\end{table*}

\subsection{Evolution of SFMC Particles and Supernova Feedback}\label{sec-III:2-TF}

In our simulation, the evolution of a SFMC particle of mass $M_{\rm MC}^{\rm init}$ describes the population of stars in it as follows:
\begin{itemize}
\item[{\it (a)}] At the birth of a SFMC particle, 24\% of $M_{\rm MC}^{\rm init}$ is instantly turned into stars.
The rest, $0.76\,\,M_{\rm MC}^{\rm init}$, is considered as atomic and molecular gas which does not participate in star formation, modeling the inefficient star formation in molecular clouds observed in local clusters \citep[e.g.][and references therein]{2007ApJ...654..304K} and predicted by numerical studies \citep[e.g.][]{2010ApJ...709..191M}.  
\item[{\it (b)}] Assuming a Salpeter initial mass function \citep[IMF;][]{1955ApJ...121..161S} between 0.1 and $300\,\,M_{\odot}$,\footnotemark\, 17\% of stellar mass, i.e. $0.04\,\,M_{\rm MC}^{\rm init}$, is massive enough to end their lives with Type II supernova explosion that peaks at the particle age of 5 Myr.  
The rest of the stellar mass, $0.2\,\,M_{\rm MC}^{\rm init}$, is considered as {\it long-lived} stellar mass, $M_{*}$. 
\item[{\it (c)}] At the end of a lifetime of a SFMC particle, $0.76+0.04 = 0.8\,\,M_{\rm MC}^{\rm init}$ is blown back into the ISM with the energy of exploding supernovae.   
\footnotetext{Our choice of an IMF reaching 300 $M_{\odot}$ instead of the more usual choice of 120 $M_{\odot}$ is motivated by recent observations suggesting that the IMF is not truncated at 100-150 $M_{\odot}$  \citep{2010MNRAS.408..731C}, as had been claimed based on earlier observations.}
\end{itemize}

Overall, the mass of a SFMC particle is expressed as 
\begin{eqnarray}
M_{\rm MC}(t) &=& M_{\rm MC}^{\rm init}  \left(1 -  0.8 \int_0^{T} \sqrt{8 \over \pi} \,\, e^{-8(T'-5)^2}\,dT' \right) \\
&=& M_{\rm MC}^{\rm init}   \left(1 - 0.8 \left[0.5 - 0.5\,\,{\rm erf} \{ \sqrt{8}(5 - T) \}  \right] \right),
\label{eq:SFMC_mass}
\end{eqnarray}
where ${\tt erf}$() is the Gauss error function, and $T = t-t_{\rm cr}$ is the particle age in Myr with particle creation time $t_{\rm cr}$.  
This formulation replicates the gradual ejection  by supernovae of 80\% of $M_{\rm MC}^{\rm init}$ that does not end up locked into stars (see Figure \ref{fig:MC_mass_description}).  
This mass, 0.8\,$M_{\rm MC}^{\rm init}$, along with $7.5 \times 10^{-7}$ of the rest mass energy of $M_{*}(T=6\, {\rm Myr}) = 0.2\,M_{\rm MC}^{\rm init}$ is returned to the cell in which the particle resides (see the right panel of Figure \ref{fig:RTF}).\footnotemark\,\,
The rate of such mass and energy ejection by supernova is proportional to $e^{-8(T-5)^2}$ which peaks at $T=5$ Myr with a width of $\sim$ 1 Myr.   
Lastly, 2\% of the ejected mass is considered as metals, contributing to the ISM enrichment. 
\footnotetext{$10^{51}$ ergs per every 750 $M_{\odot}$ of long-lived stellar mass formed.  We adopt this rather low value to make supernovae marginally effective in halting the runaway collapse of star-forming gas.  This choice is designed to further contrast the effect of stellar radiation when two fiducial simulations, MC-TF and MC-RTF, are compared.}

To summarize, we emphasize that our model incorporates three key observations of star formation in GMCs: {\it (a)} the slow and inefficient star formation in molecular clouds, {\it (b)} the commencing of Type II supernova at the death of massive stars, and {\it (c)} the recycling of a large fraction of the cloud gas, blown back into the ISM.

\subsection{Ionizing Radiation Feedback}\label{sec-III:2-RF}

Heating by photoionizing radiation from young massive stars markedly alters the environment of the stellar nursery, and it has long been postulated as one of the major drivers of self-regulated galactic star formation \citep{1979MNRAS.186...59W, 1997ApJ...476..166W, 2002ApJ...566..302M}.
In this work, we perform an explicit, three dimensional transport calculation of ionizing radiation from SFMC particles in order to evolve the radiation fields throughout the galactic ISM .   
While the radiative transfer machinery is well described in previous work \citep[e.g.][]{2002MNRAS.330L..53A, 2008ApJ...685...40W, 2011MNRAS.414.3458W, 2011ApJ...738...54K}, we briefly describe the basic details relevant to this study.  

Once a SFMC particle is created, ionizing radiation luminosity from the particle $i$ is assigned by
\begin{eqnarray}
L_{\rm MC}(i,t) = q_{\rm MC}  \, E_{\rm ph}  \, M_{\rm MC}(i, t)  
\label{eq:lum_MC}
\end{eqnarray}
where $q_{\rm MC} = 6.3 \times 10^{46} \,\,{\rm photons} \,\, {\rm s}^{-1} M^{-1}_{\odot}$ is the lifetime-averaged ionizing luminosity per solar mass in clusters \citep[Eq.(10) of][]{2010ApJ...709..424M},  $E_{\rm ph} = 16.0\, {\rm eV}$ is the mean energy per deposited photon \citep{2004ApJ...610...14W, 2006ApJS..162..281W}  fixed for monochromatic rays, and $M_{\rm MC}(i,t)$ is the SFMC particle mass as a function of particle age (see Figure  \ref{fig:MC_mass_description}).  
Readers should note that we choose to employ $M_{\rm MC}(i,t)$ rather than the stellar mass inside it. 
Our choice gives the effective ionizing luminosity per stellar mass of $1.9-2.6 \times 10^{47} \,\,{\rm photons} \,\, {\rm s}^{-1} M^{-1}_{\odot}$ at the birth of a SFMC particle, depending on the uncertainty of the lower limit of the Salpeter IMF ($0.1-1.0\,\,M_{\odot}$).  
While the value is slightly larger than what was adopted in some previous studies \citep[e.g. $\sim 10^{47} \,\,{\rm photons} \,\, {\rm s}^{-1} M^{-1}_{\odot}$ in ][]{2003ApJ...599...50F, 2012MNRAS.424..377D}, it helps to approximately encapsulate various other forms of feedback beyond photoionization, such as protostellar outflows \citep{2008ApJ...687..354N} and stellar winds \citep{2001dge..conf..181O}.\footnotemark
\footnotetext{We also note that \cite{2012MNRAS.424..377D} find that factor of two uncertainties in ionizing luminosity are less crucial in determining the amount of ionized gas than the gravitational potential in star-forming clouds.   
We will come back to this issue in \S\ref{sec-III:5-time}.}

The luminosities $L_{\rm MC}(i,t)$ are assigned only to SFMC particles that are more than 2.5 kpc away from the galactic center. 
This is to exclude the densest portion of the stellar disk which could be the result of an incorrectly structured central mass concentration.
This way, we can also concentrate on the evolution of SFMC particles in the galactic spiral arms and outer disk (see Appendix \ref{sec:appendix-A} for more discussion).
If the galacto-centric distance to the particle is greater than 2.5 kpc, then $12\times4^3$ rays ({\it Healpix} level 3) are isotropically cast from the particle.
Each ray is traced until most of its photons are absorbed or until it reaches the edge of the computational domain, while being adaptively split into child rays whenever the angular resolution associated with it grows larger than a threshold.\footnote{A ray is split as soon as the area resolved by the ray becomes larger than $0.2\, (\Delta x)^2$ of a local cell.  However, to reduce the cost of radiation calculation the ray is no longer split when it is more than 5 kpc away from the source, a distance far enough for a dwarf-sized galaxy.  In another attempt to speed up the calculation two rays may merge to form a single ray.  Merging of the rays occurs when the distance from the emitting source is more than 10 times the separation between the two sources.}
Here, achieving high resolution around SFMC particles is critical in order to adequately describe the capture of stellar radiation energy.   

From $T = 0$ to 6 Myrs after the birth of a SFMC particle, the emitted photons interact with the interstellar gas (see the middle panel of Figure \ref{fig:RTF}).  
This ionizing radiation feedback covers the period between the birth of a SFMC particle and the Type II supernova explosion by massive stars in the cluster \citep{2011MNRAS.417..950H, 2012arXiv1208.0002S}.
Though both hydrogen and helium are tracked in the chemistry module, photons interact only with hydrogen in the ISM via following three mechanisms.
First, a ray loses its photons when it ionizes hydrogen at the {\it photoionization} rate of
\begin{eqnarray}
k_{\rm ph, H} &= { P_{\,\,\rm in} (1-e^{-\tau_{\rm H}}) \over n_{\rm H} (\Delta x)^3 dt_{\rm ph} }
\end{eqnarray}
where $P_{\,\,\rm in}$ is the number of incoming photons, $\tau_{\rm H} = n_{\rm H} \sigma_{\rm H} dl$ is the optical depth of a cell,  $n_{\rm H}$ is the neutral hydrogen number density, $\sigma_{\rm H}$ is the energy-dependent hydrogen photoionization cross-section \citep{1996ApJ...465..487V}, and $dl$ is the path length through the cell.  
The radiation timestep $dt_{\rm ph}$ set by the code is typically comparable to the finest hydrodynamical timestep.
Second, the excess energy above the ionization threshold heats the cell with the {\it photoheating} rate of
\begin{eqnarray}
\Gamma_{\rm H} &=& k_{\rm ph, H} (E_{\rm ph} - E_{i, \rm H})  \nonumber \\
&=& { P_{\,\,\rm in} (1-e^{-\tau_{\rm H}}) (E_{\rm ph} - E_{i, \rm H}) \over n_{\rm H} (\Delta x)^3 dt_{\rm ph} }
\end{eqnarray}
where $E_{i, \rm H} = 13.6\, {\rm eV}$ is the ionization threshold.  
Third, when photons are absorbed by the gas cell, the added acceleration by {\it radiation pressure} onto the cell is calculated as
\begin{eqnarray}
{\bold a}_{\rm ph} = {d{\bold p}_{\rm ph} \over m_{\rm cell} dt_{\rm ph}} &=& { P_{\,\,\rm lost} (E_{\rm ph}/c) \over m_{\rm cell} dt_{\rm ph} } {\hat {\bold r}} \nonumber \\ 
&=& { P_{\,\,\rm in} (1-e^{-\tau_{\rm H}}) E_{\rm ph} \over \rho_{\rm gas} (\Delta x)^3 c dt_{\rm ph} } {\hat {\bold r}}
\end{eqnarray}
where $d{\bold p}_{\rm ph}$ is the photon momentum exerted onto the cell in $dt_{\rm ph}$, $P_{\,\,\rm lost}$ is the number of photons lost in the cell, and ${\hat {\bold r}}$ is the directional unit vector of the ray away from the source.

It is worth briefly noting a few points on stellar radiation feedback before we proceed:
{\it (a)} Only {\it hydrogen} photoionization and photoheating are taken into account in the reported simulation and subsequent analysis. 
{\it (b)} Dust extinction is not included in our calculation.   
This would not significantly change our escape fraction analysis in \S\ref{sec-III:5}, as Figure 1 and \S4 of \cite{2008ApJ...672..765G} demonstrates that the absorption by dust does not substantially alter the escape fraction of ionizing radiation above the Lyman limit (see Appendix \ref{sec:appendix-B} for more discussion).  
{\it (c)} By the same token, infrared photons re-emitted by UV-irradiated dust grains and the momentum imparted by them are not explicitly considered in our calculation. 
Partly because of this reason, radiation pressure purely by the hydrogen-ionizing photons does not play a major role in reported results.
Including the trapped radiation pressure by these absorbed and re-scattered IR photons might have enhanced the negative feedback effect \citep[e.g.][]{2011MNRAS.417..950H}, although recent radiation hydrodynamics simulations \citep[e.g.][]{2012ApJ...760..155K, 2013arXiv1302.4440K} suggest that such an enhancement due to trapping is significantly less than previously assumed.

Table \ref{table-III:desc} summarizes the feedback models we employ in the study.  
In MC-TF, only supernova feedback is included.   
In MC-RTF, both channels of stellar feedback, ionizing radiation and supernova explosion, are simultaneously considered.

\begin{figure*}[t]
\epsscale{1.05}
\plotone{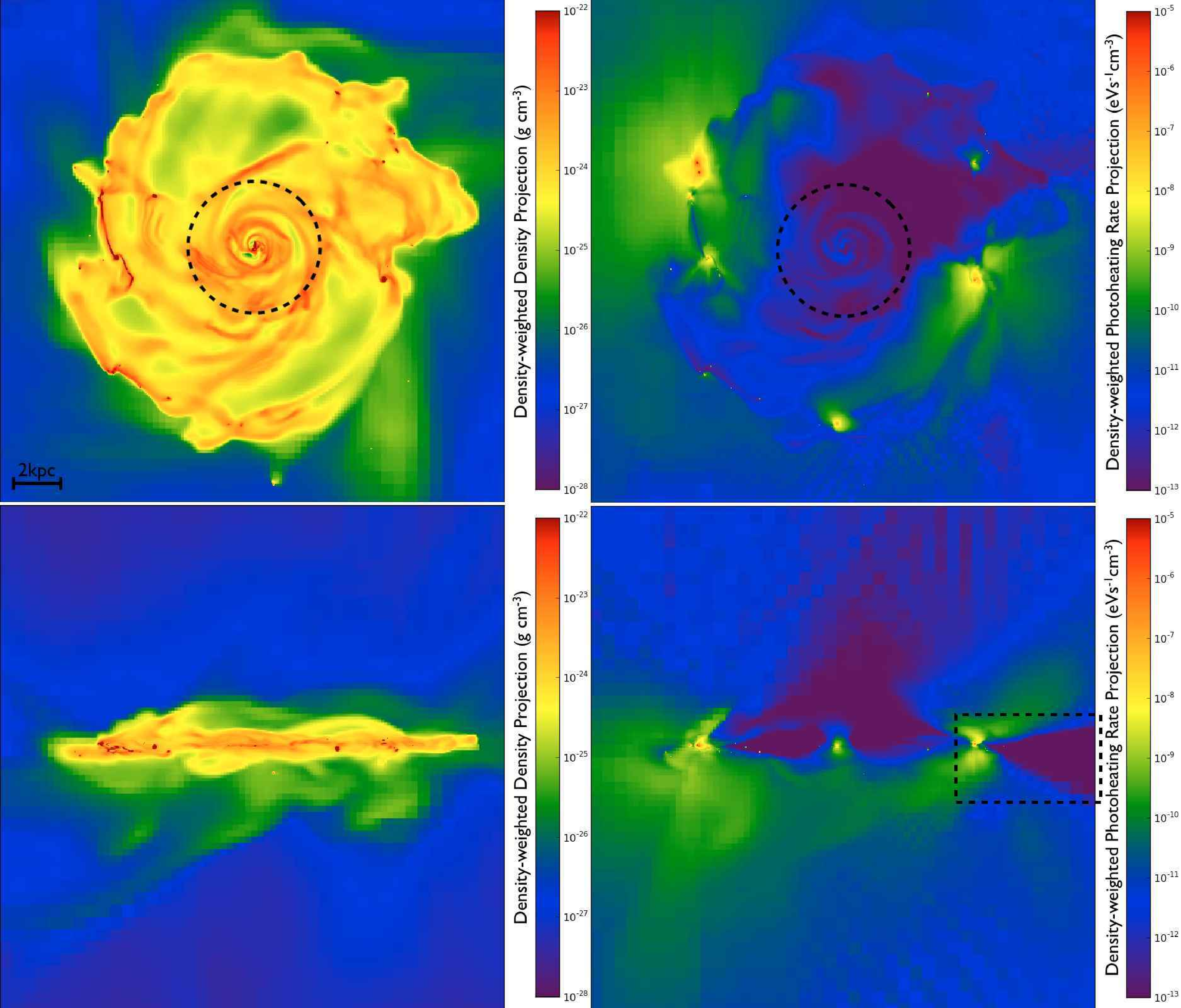}
    \caption{{\it Left columns:} density-weighted projection of density in the MC-RTF run in a 20 kpc box at 13.3 Myr into the high-resolution evolution (top: face-on / bottom: edge-on).  High-density clumps which bear a number of SFMC particles are noticeable.  
 {\it Right columns:} density-weighted projection of photoheating rates, in which the shaded region along the right side of the galactic plane is pronounced (black dotted box).  Note that the photoheating rates are low in the galactic center ($<$ 2.5 kpc in radius, dotted circles) because the particles there are not ray-emitting.
\label{fig:gammaHI_eVscm3}}
\end{figure*}

\section{Initial Conditions}\label{sec-III:3}

We simulate the evolution of SFMC particles and their embedding dwarf-sized galactic disk at 3.8 pc resolution.
The initial condition of our experiment is detailed in this section. 

\subsection{Data Conversion Pipeline}

The {\it Gadget-2} \citep{2005MNRAS.364.1105S} initial condition generator {\it StarScream}\footnote{The authors are grateful to Jay Billings for making his code public and available for this study:  http://code.google.com/p/starscream/.}  is used to create an $N$-body dataset of an idealized isolated galaxy, to which we add gas particles by splitting collisionless particles.  
Then, the {\it Gadget}-to-{\it Enzo} converter {\it hullMethod} employs IDL functions {\tt qhull} (Delaunay tessellation) and {\tt qgrid3} (linear interpolation) to map the particle data onto an adaptively refined, oct-tree structured mesh.
For details of the data conversion pipeline we refer the readers to \cite{2008AIPC..990..429K, 2009ApJ...694L.123K}.

\subsection{Setting Up A Dwarf-sized Galactic Disk}

We construct a dark matter halo of $2.3\times10^{11} M_{\odot}$ that follows the Navarro-Frenk-White profile with a concentration parameter $c = 10$ \citep{1997ApJ...490..493N}.  
Gas grids are generated by splitting the particles with an initial baryonic fraction of 10\%, and with initial metallicity $Z_{\rm init} = 0.003 \,\,Z_{\odot}$ everywhere.  
In addition, a collective rotation of spin parameter $\lambda = 0.05$ is provided to both gas and dark matter.  
The value of $\lambda$ is chosen to be slightly higher than the cosmological average found in simulations \citep{2001ApJ...555..240B} so that it helps to reliably create a rotationally supported galactic disk.
We then produce a star-forming disk galaxy embedded in a dark matter halo by performing a coarse resolution (15.2 pc) simulation for $\sim$ 1 Gyr in a $32^3$ root grid box of $1\,{\rm Mpc}^3$. 
While the galaxy itself occupies only a small portion of this root level box, the large box size is selected so that any boundary effect is marginalized.
When a relaxed, well-defined galactic disk has emerged, $\sim 1.3\times10^{10} M_{\odot}$ is in $3.4\times10^{6}$ star cluster particles formed with a coarse-resolution refinement strategy and SFMC particle formation recipe.  
$\sim 6.5\times10^{9} M_{\odot}$ is in gaseous form either in the ISM or in the halo.
These are all embedded in a halo of $1.3\times 10^7$ dark matter particles. 

We now employ high-resolution (3.8 pc) refinement criteria to resolve the galaxy down to the size of an individual molecular cloud, and make the entire galaxy evolve for another $\sim$ 30 Myrs.  
Such high resolution is obtained by applying 2 additional levels of refinement along with the formation criteria of $\sim$ 1000 $M_{\odot}$ SFMC particles (see \S 2.2).
By applying high resolution refinement only well into the galactic evolution, one can save the computational expense of performing an extremely high resolution calculation for a galactic dynamical time ($>$ 300 Myrs), but can still observe the behavior of star-forming gas clumps for their typical lifetime ($\sim$ 30 Myrs).

\section{Results I: Overview} \label{sec-III:4}

A suite of simulations with different stellar feedback is performed to examine the behavior of SFMC particles and their embedding galaxy in one self-consistent framework at high spatial resolution.
We first overview the performed runs.

\subsection{Summary of Performed Runs}  \label{sec-III:4-sum}

Each of the calculations is carried out on 16 to 32 processors of the Ranger cluster\footnote{Sun Constellation Linux Cluster, Infiniband-connected AMD, 16 cores per node, 2 GB memory per core} at the Texas Advanced Computing Center at the University of Texas at Austin.   
In a typical simulation, a galaxy is resolved with $\sim 2.0\times10^7$ total computational elements ($\sim 1.7\times10^7$ particles and $\sim 160^3$ cells in $\sim 10^3$ grids).  
In a simulation with both channels of stellar feedback (MC-RTF in Table \ref{table-III:desc}), there are about 3000 to 10000 active SFMC particles that participate in radiative transfer calculation. 
It takes $\sim$ 2000 CPU hours on average to evolve this system for 1 Myr. 
A portion of the post-production procedure and data analysis is conducted on the Orange cluster at the Kavli Institute for Particle Astrophysics and Cosmology at Stanford University.   

\subsection{Self-regulated Star Formation}  \label{sec-III:4-SF}

In order to exclude the period in which the galaxy is trying to reach a new equilibrium with freshly imposed high resolution, we investigate the galaxy from 10 to 30 Myr into the evolution in the subsequent analysis.
During this period of $\sim$ 20 Myrs, the run with both stellar radiation and supernova feedback forms 22.4\% less stellar mass (MC-RTF, $3.57\,\,M_{\odot}{\rm yr}^{-1}$ on average) than the run with only supernova feedback (MC-TF, $4.60\,\,M_{\odot}{\rm yr}^{-1}$).
Our realistic description of stellar feedback helps to self-regulate galactic star formation, and provide a reasonable star formation rate during the analyzed period of the simulation.  
The suppression ratio could have been larger had the SFMC particles in the inner disk also radiated ionizing photons ($<$ 2.5 kpc from the galactic center; see \S\ref{sec-III:2-RF}).  
Star formation relations in connection with the local Kennicutt-Schmidt relation \citep{2007ApJ...671..333K, 2008AJ....136.2846B} will be extensively discussed in an accompanying future study \citep{2012arXiv1210.6988K}.  
For more discussion on self-regulated star formation relations, we refer interested readers to this paper.   

\begin{figure*}[t]
\epsscale{1.2}
\plotone{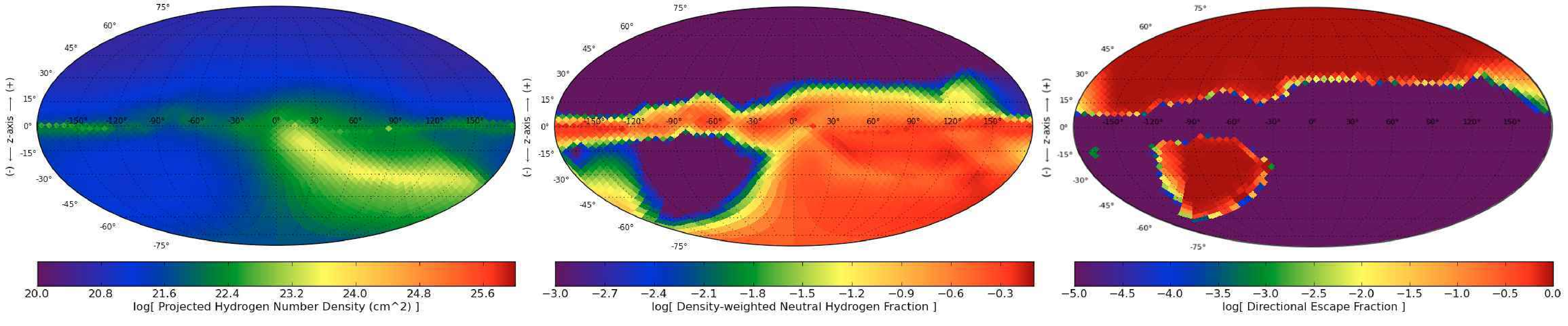}
    \caption{{\it Left:} all-sky map of projected neutral hydrogen number density (Mollweide equal-area projection) from a radiating SFMC particle in the MC-RTF run, seen at 100 kpc at 29.9 Myr into the high-resolution evolution.  It displays two notable structures: the galactic disk across the mid-plane, and the gaseous clump close to this particle in the bottom right.  
{\it Middle:} all-sky map of density-weighted average of neutral hydrogen fraction seen at 100 kpc.  
{\it Right:} all-sky map of photon escape fraction at 100 kpc along different lines of sight, $f_{\rm esc}(i, \theta, \varphi)$.  Ionizing photons escape neither from the galactic plane nor from the nearby gas clump because of their high column densities of neutral hydrogen.
\label{fig:healpix_example}}
\end{figure*}

\section{Results II: Escape of Ionizing Photons} \label{sec-III:5}

As a first application of our sophisticated feedback model, we investigate the escape of ionizing stellar radiation in the MC-RTF run in which both ionizing radiation and supernova feedback are considered (see Table \ref{table-III:desc}).  
We start this section by reviewing the recent theoretical advances made in the field.   

\subsection{Theoretical Background}  \label{sec-III:5-allsky}

Hydrogen ionizing radiation from dwarf-sized galaxies is thought to be an important driver of reionization in the early universe \citep[e.g.][]{1999ApJ...514..648M, 2008ApJ...688...85F, 2010MNRAS.409..855B, 2012MNRAS.423..862K}. 
Its contribution may have increased in time since $z \sim 6$, relative to that of quasars and other sources.  
Also, galactic ultraviolet radiation filtered through the intergalactic medium (IGM) is a major source of the metagalactic ultraviolet background.
This background radiation reduces the gas accretion onto less massive halos and evaporates their existing gas \citep[e.g.][]{1996ApJ...461...20H, 2001cghr.confE..64H, 2011MNRAS.412.2543C}.   

As such, escape of ionizing radiation from a dwarf galaxy has been a fiercely researched topic in extragalactic astronomy.  
However, the estimates of galactic escape fraction vary widely, indicating the difficulty to comprehend the behavior of ionizing photons within a morphologically complicated galactic structure. 
Various observational studies place the galactic escape fraction at a few percent for low-redshift \citep[e.g.][]{2001ApJ...558...56H, 2006A&A...448..513B, 2012ApJ...755...40P} and high-redshift galaxies \citep[e.g.][]{2002ApJ...568L...9G, 2006ApJ...651..688S, 2009ApJ...692.1287I, 2011ApJ...736...41B}, with suggested decrease in time likely due to more clumpy structures in more evolved galaxies \citep[e.g.][]{2006MNRAS.371L...1I}.

Theoretical investigations have also produced a similar value for galactic escape fraction. 
Analytic estimations employ idealized galaxy models and/or Monte Carlo radiative transport calculation to evaluate the photon escape fraction with varying galactic parameters \cite[e.g.][]{2000ApJ...545...86W, 2002MNRAS.337.1299C, 2011ApJ...731...20F}. 
Other authors have acquired the escape fraction by separately post-processing a simulated galaxy which did not integrate the effect of ionizing stellar radiation in the equation of hydrodynamics \citep[e.g.][]{2003ApJ...599...50F, 2007ApJ...668..674R, 2010ApJ...710.1239R, 2009MNRAS.398..715Y, 2011MNRAS.412..411Y, 2011A&A...530A..87P}.
While self-consistent radiation hydrodynamics simulations have been used to study reionization and galaxy formation at low resolution \cite[e.g.][]{2011MNRAS.412..935P, 2012arXiv1209.4143H}, only occasionally have they been adopted to probe galactic escape fraction with high spatial resolution \citep[e.g.][]{2008ApJ...672..765G, 2009ApJ...693..984W}.

As described in \S\ref{sec-III:2}, our work incorporates hydrodynamics, chemistry, star formation, and ionizing radiation from star-forming clouds in one self-consistent framework with an unprecedented resolution.  
We thereby do not neglect the hydrodynamic and chemical influence of the radiation on the ISM.  
Compared to previous studies, the ray-emitting sources in our simulations represent $\lesssim$10000 radiating SFMC particles spatially resolved in 3.8 pc resolution, scattered on a galactic disk. 
Evolutionary stages during the lifetime of the molecular clouds are also resolved by the corresponding temporal resolution (see \S\ref{sec-III:2-TF} and \S\ref{sec-III:2-RF}).  
Additionally, our study explores a more massive and well-defined galactic disk than the $3 \times 10^6 \,\,\,{\rm to}\,\,\, 3 \times 10^9 \,M_{\odot}$ halos considered in \cite{2009ApJ...693..984W}, therefore complementing their work. 
Last but not least, our newly-developed approach to post-process the simulated data facilitates groundbreaking studies such as examining the evolution of escape fraction {\it per star-forming clump}.  
In the remainder of \S\ref{sec-III:5}, we focus on the escape of ionizing photons and its angular, temporal, and spatial variations in the MC-RTF run, unless stated otherwise.  

\begin{figure*}[t]
\epsscale{1.17}
\plotone{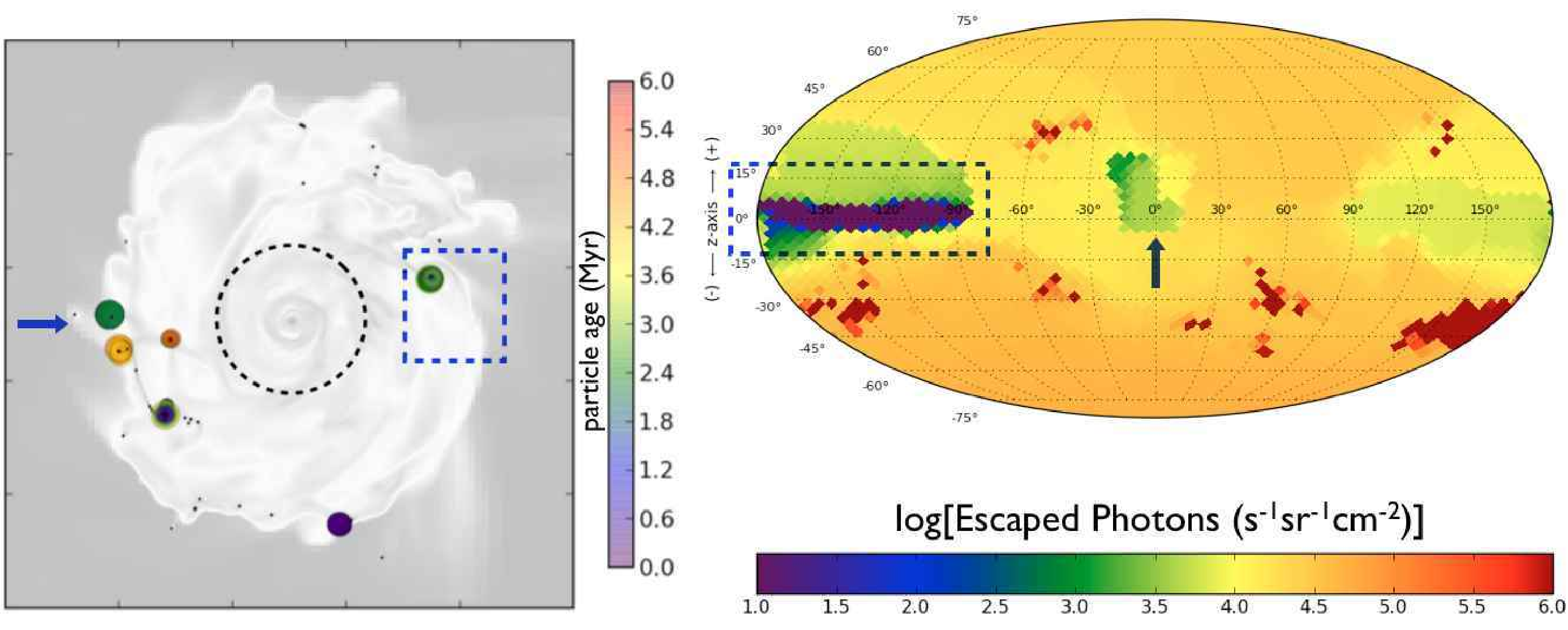}
    \caption{{\it Left:} the locations of SFMC particles sized by their escape fractions, $f_{\rm esc}(i)$, and colored by their ages.  $f_{\rm esc}(i)$ values are evaluated at 30.7 Myr into the high-resolution evolution, and plotted on top of a black-and-white face-on projection of gas density in a 20 kpc box.  A single-pixel dot means a SFMC particle with zero escape fraction.   
    {\it Right:} the weighted sum of the fluxes of escaped photon numbers from SFMC particles along different lines of sight, and seen at 100 kpc from the ray-emitting sources.  The latitude corresponds to the direction vertical to the galactic disk plane, while $0^{\circ}$ in the longitudinal coordinate correlates with the left side of the galactic plane in the left panel (blue arrow in each panel).  The average escape fraction at this snapshot is $f_{\rm esc} = 0.56 \%$.
\label{fig:fesc_clump_allsky}}
\end{figure*}

\subsection{Escape Fraction via All-Sky Projection}  \label{sec-III:5-allsky}

Figure \ref{fig:gammaHI_eVscm3} compiles the projected density and photoheating rates in the MC-RTF run at 13.3 Myr into the evolution with high resolution.  
It is straightforward to locate the actively radiating SFMC particles around which the photoheating rates peak.   
Also noticeable is the preferential escape of ionizing radiation towards polar directions.  
In particular, the shaded region along the right side of the galactic disk plane is highly pronounced in the bottom right image of the Figure \ref{fig:gammaHI_eVscm3} (black dotted box). 
Note that the photoheating rates are relatively low in the galactic center because the particles there are not emitting radiation (black dotted circles, $<$ 2.5 kpc from the galactic center; see \S\ref{sec-III:2-RF}).
Radiation fields such as photoheating rates (summed for all incoming photons to the cell) are tracked by the radiation solver in each computational cell, and later used in chemistry and cooling calculations. 

We, however, aim to evaluate the fraction of ionizing photons that escapes from an individual star-forming clump, or from a galaxy.
While informative, {\it integrated} radiation fields such as in Figure \ref{fig:gammaHI_eVscm3} are less illuminating for our purpose.
To compute the escape fraction of photons from an individual SFMC particle, all-sky maps of the following quantities centered on a SFMC particle are generated and shown in Figure \ref{fig:healpix_example}:  {\it (a)} the column density of the number of neutral hydrogen, $\Sigma_{n_{\rm H}, 100\,\, {\rm kpc}}$, {\it (b)} the density-weighted average of neutral hydrogen fraction, and {\it (c)} the photon escape fraction along different lines of sight.  
Each of these maps is seen at 100 kpc away from the ray-emitting source, and at 29.9 Myr into the high-resolution evolution.
The angular resolution of these maps, $N_{\rm res} = 12 \times 4^4 = 12 \times 16^2$ = 3072, corresponds to the ${\it Healpix}\,\,\, {\rm level} = 4$ or $N_{\rm side} = 16$.
The leftmost panel of the projected number density shows two prominent gas structures: the galactic disk across the mid-plane, and the gaseous clump in the bottom right of the map that happens to be very close to this particular source.  

Most notably, the rightmost panel of Figure \ref{fig:healpix_example} illustrates the all-sky map of escape fraction at 100 kpc along different lines of sight from the particle.  That is
\begin{eqnarray}
f_{\rm esc}(i, \theta, \varphi) = e^{-\tau_{\rm H, 100\,\, kpc} (i, \theta, \varphi)}
\label{eq:directional_escape_fraction}
\end{eqnarray}
where $i$ is the source identification number, $\theta$ and $\varphi$ are the angles to the pixel, and $\tau_{\rm H, 100\,\, kpc} = \sigma_{\rm H} \Sigma_{n_{\rm H}, 100\,\, {\rm kpc}}$ is the optical depth at 100 kpc.
It is evident that ionizing photons are prohibited from escaping the galactic plane or the nearby gas clump due to their high column densities.
Because the initial rays are cast isotropically (see \S\ref{sec-III:2-RF}), the overall escape fraction of ionizing photons from this particular source is easily obtained by averaging $f_{\rm esc}(i, \theta, \varphi)$ over the entire sky, or equivalently, over $N_{\rm res}$ equal-area pixels.  
Hereafter the escape fraction {\it per SFMC particle}, $f_{\rm esc} (i)$, is defined as such; in other words, a fraction of photons that reached a 100 kpc sphere without being absorbed:
\begin{eqnarray}
f_{\rm esc}(i) \,\, &=& \,\, {1\over 4\pi} \int_{0}^{2\pi} \int_{0}^{\pi} f_{\rm esc}(i, \theta, \varphi) \, {\rm sin}\theta d\theta d\varphi \\
&\approx& {1 \over N_{\rm res}} \displaystyle\sum_{\theta, \varphi}  f_{\rm esc}(i, \theta, \varphi).
\end{eqnarray}
Further, the {\it average} escape fraction, $f_{\rm esc}$, is defined as the mean of $f_{\rm esc} (i)$ over the entire set of ray-emitting sources considered weighted by the initial photon number from each source, $q_{\rm MC}  \, M_{\rm MC}(i)$ in Eq.(\ref{eq:lum_MC}).  This could be regarded as the escape fraction {\it of a galaxy} (but see Appendix \ref{sec:appendix-A}):
\begin{eqnarray}
f_{\rm esc} \,\, = \,\, { \displaystyle\sum_{i} \,\, q_{\rm MC}  \, M_{\rm MC}(i) \, f_{\rm esc}(i) \over \displaystyle\sum_{i} \,\, q_{\rm MC}  \, M_{\rm MC}(i) }.
\end{eqnarray}

Note that simply by integrating the neutral hydrogen number density along different lines of sight we have acquired $\tau_{\rm H, 100\,\, kpc}(i, \theta, \varphi)$, and hence $f_{\rm esc}(i)$ and $f_{\rm esc}$.  
Our post-production procedure does not include any additional radiative transfer, and is interactive and highly user-adaptive.\footnotemark\,
However, while our calculation is done in post-processing, we demonstrate in \S\ref{sec-III:5-comp} that the result we obtain depends critically on including the photoionizing radiation self-consistently during the main calculation.
\footnotetext{Alternatively one may explicitly count the number of photons that have been traced for more than 100 kpc, within the radiative transfer simulation.  This type of calculation is non-interactive and may add additional computational cost.}

\begin{figure*}[t]
\epsscale{1.2}
\plotone{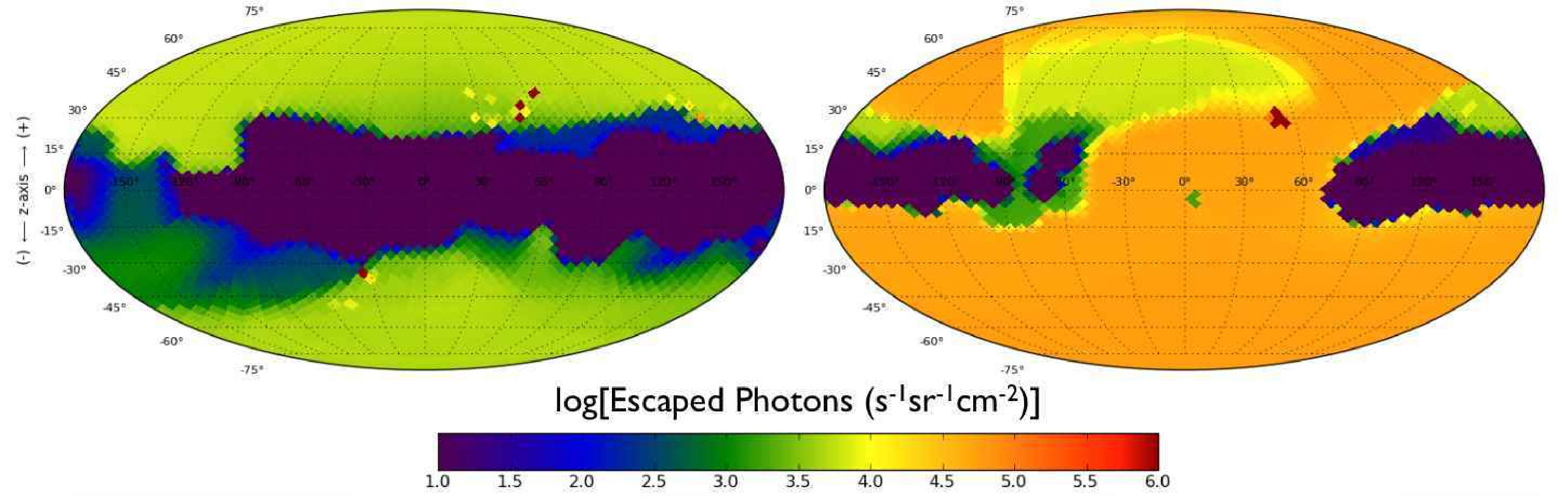}
    \caption{The escaped ionizing photon fluxes measured at 17.7 ({\it left}) and 19.6 Myr ({\it right}) into the high-resolution evolution.  For each snapshot, the average escape fraction is $f_{\rm esc} = $ 0.08\% and 1.1\%, respectively.  Note that at 17.7 Myr ionizing photons rarely escape along the disk plane of the simulated galaxy.  See the caption of Figure \ref{fig:fesc_clump_allsky} for a detailed description of the coordinate system.  
\label{fig:fesc_compare_allsky}}
\end{figure*}

\subsection{Escaped Ionizing Photon Flux from A Galaxy and Its Angular Variation}  \label{sec-III:5-angle}

Displayed on the left panel of Figure \ref{fig:fesc_clump_allsky} are the locations of SFMC particles whose marker sizes are proportional to the escape fractions from the particles, $f_{\rm esc}(i)$,  at 30.7 Myr into the high-resolution evolution.  
The panel depicts the major contributors of high $f_{\rm esc}(i)$ to the overall averaged escape fraction, $f_{\rm esc} = 0.56 \%$.   
The age of each particle is represented by the color of the circle.  
Note that the inner disk is not populated by radiating SFMC particles because the particles there are not ray-emitting (black dotted circle, $<$ 2.5 kpc from the galactic center; see \S\ref{sec-III:2-RF} and  Appendix \ref{sec:appendix-A}). 
The average escape fraction is heavily dominated by only a small number of SFMC particles with high $f_{\rm esc}(i)$, while the majority of SFMC particles have negligible escape fraction as shown by points with a single pixel.  
We will come back to this in \S\ref{sec-III:5-space} for more discussion.  

For now we concentrate primarily on the average escape fraction and the total flux of escaped photons from all the sources considered on the galactic spiral arms (excluding the central 2.5 kpc).  
The right panel of Figure \ref{fig:fesc_clump_allsky} demonstrates the weighted sum of the fluxes of escaped photon numbers from all SFMC particles along different lines of sight, measured at 100 kpc from the ray-emitting sources. 
This could be interpreted as a escaped photon flux from the galaxy, assuming that 100 kpc distance is far enough to ignore the separation among ray-emitting SFMC particles.
Note that $0^{\circ}$ in the longitudinal coordinate corresponds to the left side of the galactic disk plane in the left panel (blue arrow in each panel), and $\pm180^{\circ}$ to its right side. 
Therefore, the prominent shaded region along the galactic plane between $\varphi = -180^{\circ}\,\, {\rm and}\,\, -90^{\circ}$ points to the right side of the galactic disk (blue dotted box in each panel).  
As seen in the blue dotted box, the right side of the galaxy hosts only one distinct star-forming clump with many high $f_{\rm esc}(i)$ SFMC particles. 
Photons emanating from this clump are often blocked by thick gas spiraling around it.  

Throughout $\sim$ 20 Myrs of high-resolution evolution, we find that the flux of escaped photons is in general the smallest along the galactic disk plane.  
Moreover, the escape fraction varies by up to 4 to 5 orders of magnitude along different lines of sight.   
For example, Figure \ref{fig:fesc_compare_allsky} shows the escaped ionizing photon fluxes at two different epochs.   
At 17.7 Myr ionizing photons rarely escape along the disk plane of the simulated galaxy.  
At 19.6 Myr photons do escape towards the right side of the galaxy, but not through its left side.  
Zero or negligible escape fraction is found due to the large column density of neutral hydrogen in a well-defined galactic disk, which is seldom observed in prior numerical studies of cosmological galaxies \citep[e.g.][]{2008ApJ...672..765G, 2009ApJ...693..984W}.
Nonetheless, the flux of escaped photons is not strongly beamed, but rather manifests a large opening angle of more than $60^{\circ}$ from the galactic pole.   
We caution the readers that the beaming of the escaped photon flux might have been enhanced had we allowed SFMC particles in the inner part of the galactic disk to radiate.
Photon fluxes from the sources in the inner disk might have preferentially contributed towards the galactic pole due to disk geometry (but see Appendix \ref{sec:appendix-A}).  

\subsection{Temporal Variation of Average Escape Fraction} \label{sec-III:5-time}

\begin{figure}[t]
\epsscale{1.18}
\plotone{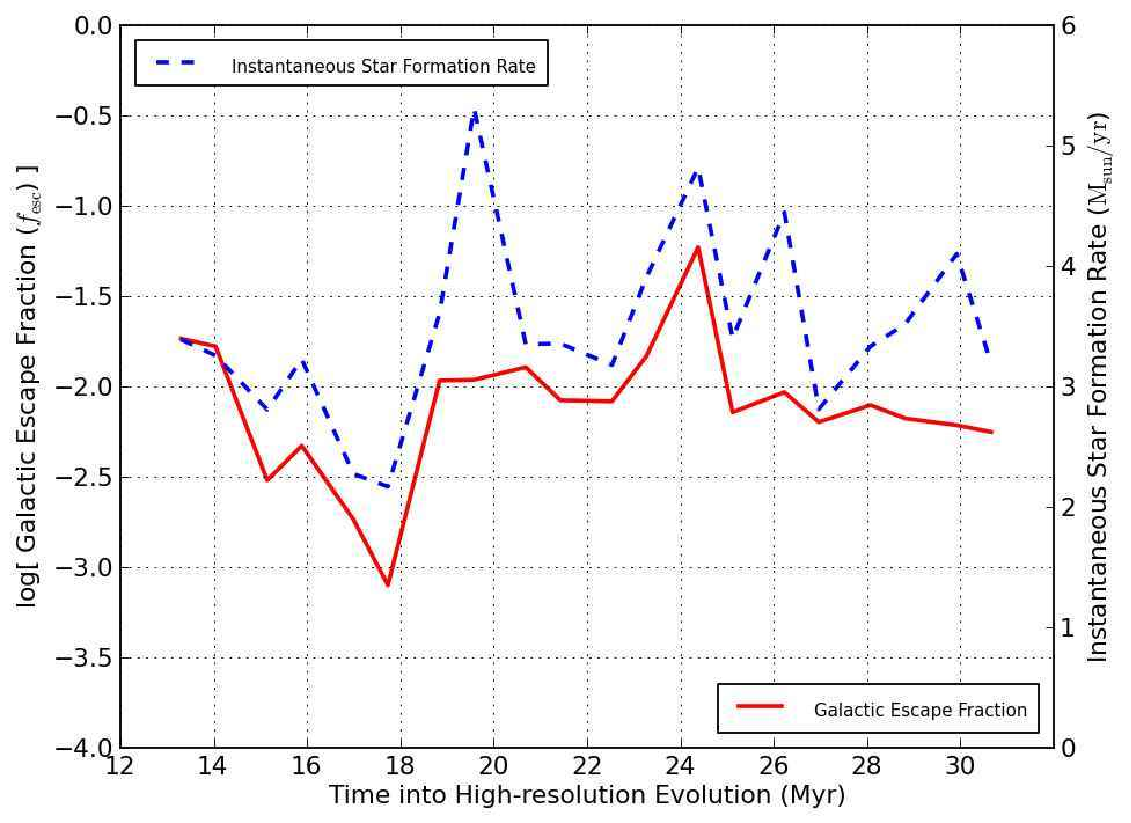}
    \caption{{\it Solid line:} time evolution of the average escape fraction from all radiating sources on the spiral arms (excluding the central 2.5 kpc), $f_{\rm esc}$, from 13.3 Myr to 30.7 Myr of the high-resolution evolution in the MC-RTF run.  It fluctuates between 0.08\% at 17.7 Myr and 5.9\% at 24.4 Myr.  
{\it Dashed line:} instantaneous star formation rate in the unit of $M_{\odot} {\rm yr}^{-1}$.
\label{fig:fesc_temporal}}
\end{figure}

As discussed in \S\ref{sec-III:5-allsky},  angularly averaged escape fractions can be obtained by integrating the escaped photon flux over the entire sky.   
Figure \ref{fig:fesc_temporal} exhibits the time evolution of the average escape fraction from all radiating sources on the spiral arms (excluding the central 2.5 kpc), $f_{\rm esc}$, from 13.3 Myr to 30.7 Myr of the high-resolution evolution.  
While the mean value of $f_{\rm esc}$ is 1.1\%, it fluctuates between 0.08\% at 17.7 Myr and 5.9\% at 24.4 Myr.    
The time-averaged $f_{\rm esc}$ of 1.1\% is lower than what was previously found in simulated low-mass dwarf galaxies at high redshift \citep[$M \lesssim 10^{9.5} M_{\odot}$;][]{2009ApJ...693..984W}, but it is in line with the values  measured in a simulation of a larger galaxy \citep[$M \gtrsim 10^{11} M_{\odot}$;][]{2008ApJ...672..765G} or an idealized galaxy with a rotationally supported disk \citep{2011A&A...530A..87P}.
Also found is a mild positive correlation between $f_{\rm esc}$ and the instantaneous star formation rate,  $dM_*/dt$ (with $dt = 0.37$ Myr), plotted with a dashed line.
The large variation of $f_{\rm esc}$ up to more than an order of magnitude in just a few Myrs might be related to the change in instantaneous star formation rate.  
These observations are consistent with the previous escape fraction studies using radiation hydrodynamics simulations \cite[e.g.][]{2008ApJ...672..765G, 2009ApJ...693..984W}.  

We, however, caution that $f_{\rm esc}$ might have changed if the SFMC particles in the inner disk had radiated ionizing photons  (see Appendix \ref{sec:appendix-A}).  
Other sources of uncertainty include: {\it (a)} the change in supernova energy that disperses the surrounding gas, {\it (b)} the change in ionizing luminosity, Eq.(\ref{eq:lum_MC}), although \cite{2012MNRAS.424..377D} argues that a factor of a few uncertainties in ionizing luminosity are not of significant importance in determining the amount of ionized gas in star-forming clouds, and {\it (c)} the inclusion of dust extinction, although it does not substantially change $f_{\rm esc}(i)$ above the Lyman limit \citep[see Appendix \ref{sec:appendix-B} and][for more discussion]{2008ApJ...672..765G}.

\begin{figure*}[t]
\epsscale{1.15}
\plotone{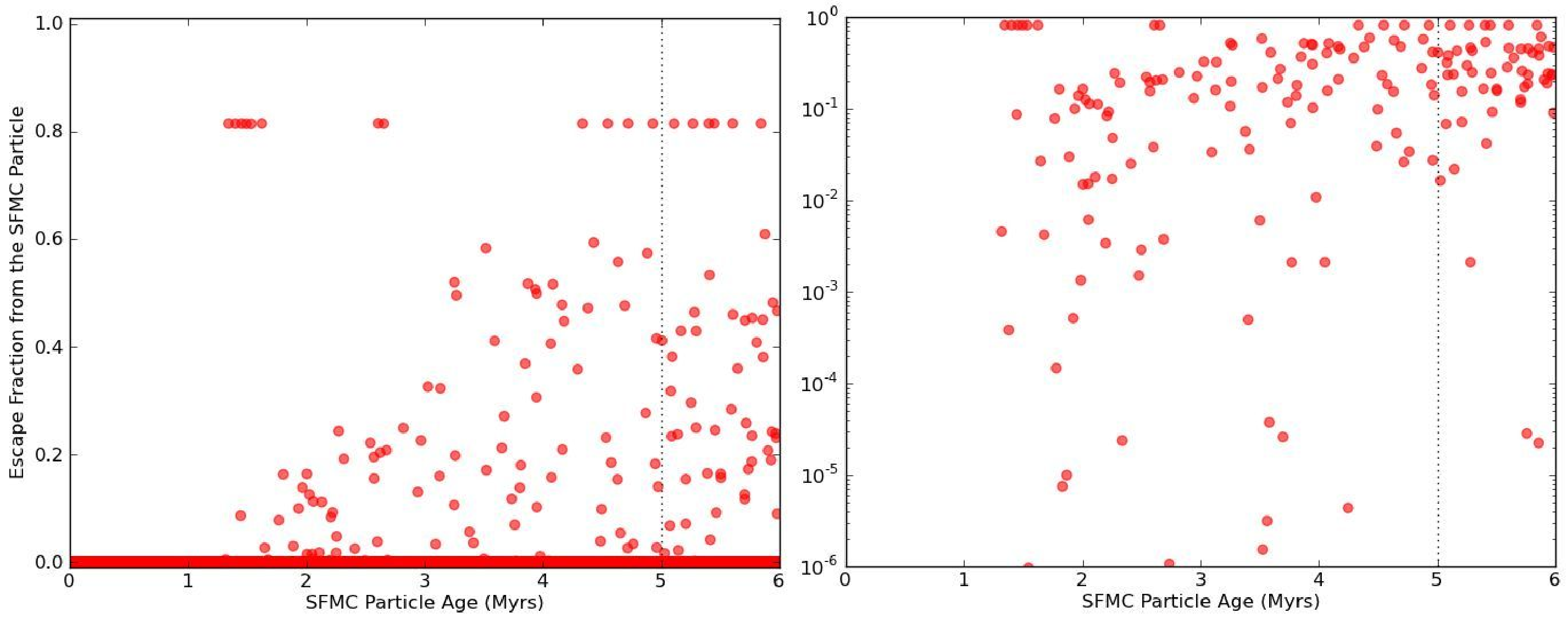}
    \caption{The distribution of actively radiating SFMC particles (i.e. particle age $T(i) < 6$ Myrs) on the plane of the particle age and the escape fraction per particle, $f_{\rm esc}(i)$, measured at 29.9 Myr into the high-resolution evolution with two types of $y$-axis, linear ({\it left}) and logarithmic ({\it right}).  At this snapshot the average escape fraction is $f_{\rm esc} = 0.61\%$.  Note that most of the SFMC particles fall on the {\it x}-axis on the left panel with negligible $f_{\rm esc}(i)$; only 1.6\% of the actively radiating particles are shown to have $f_{\rm esc}(i)$ greater than 1\%.  Note also that the escape fraction slowly rises as a function of particle age.  This trend is in part aided by the supernova explosion feedback that peaks at $T(i) = 5$ Myr, which is marked with a dotted vertical line.  
\label{fig:age_vs_fesc}}
\end{figure*}

\subsection{Escape Fraction Per SFMC Particle} \label{sec-III:5-space}

Two critical factors that distinguish our work from previous escape fraction studies are that: {\it (a)} the ray-emitting sources in our simulations model star-forming molecular clouds, and {\it (b)} our new post-production procedure enables us to measure the evolution of escape fraction {\it per SFMC particle}. 
Understanding how the ionizing stellar radiation feedback manifests itself on an evolving star-forming clump in a galactic context is of vital interest, especially in comparison with GMC-scale studies in which ionizing feedback is considered \citep[e.g.][]{2006ApJ...653..361K, 2007ApJ...671..518K, 2007ApJ...668..980M, 2010ApJ...715.1302V, 2011MNRAS.414.1747A, 2011ApJ...738..101G, 2011MNRAS.414..321D, 2012MNRAS.424..377D, 2012ApJ...745..158G}.  
Moreover, observationally, stellar radiation leaked from individual molecular clouds is thought to power the ionization of diffuse ionized gas (DIG) in a galaxy, and maintains the DIG layer above and below the galactic disk \citep[e.g.][]{1984ApJ...282..191R, 2003ApJ...586..902H, 2006ApJ...644L..29V}.  
In this section, we inspect the escape fraction per SFMC particle, $f_{\rm esc}(i)$, and how it evolves as the particle ages.  

As examined in \S\ref{sec-III:5-angle}, the average escape fraction is dominated by a small number of SFMC particles with very high $f_{\rm esc}(i)$.  
At the same time a large fraction of SFMC particles have zero or negligible escape fraction as shown in Figure \ref{fig:fesc_clump_allsky} by points with a single pixel. 
This observation is dramatically displayed in Figure \ref{fig:age_vs_fesc} where actively radiating SFMC particles are scattered on the plane of particle age and the escape fraction per particle.
This snapshot is made at 29.9 Myr into the high-resolution evolution.  
It is worth noting that out of 8709 actively radiating SFMC particles at this point only 140 have $f_{\rm esc}(i)$ greater than 1\%. 
This is merely 1.6\% of the entire population of SFMC particles.  
Among these the variation of $f_{\rm esc}(i)$ is broad, from just $\sim$ 1\% up to $\sim$ 80\%, a feature similar to what \cite{2012MNRAS.424..377D} found.
On the other hand, the majority of the SFMC particles fall on top of the {\it x}-axis on the left panel of Figure \ref{fig:age_vs_fesc} with negligible $f_{\rm esc}(i)$.
This observation strongly supports the argument that the average escape fraction is determined by a few SFMC particles with exceptionally high escape fractions \citep[e.g.][]{2008ApJ...672..765G}.  
(The following toy model may help elucidate our finding:
if galactic $f_{\rm esc}$ is 1\%, then it is more plausible that 1\% of the SFMC particles have $f_{\rm esc}(i) = 100\%$, rather than 100\% have $f_{\rm esc}(i) = 1\%$.)
From this observation, one may claim that the escape of ionizing photons is primarily determined by the small-scale properties of the star-forming clouds in the vicinity of a SFMC particle, not by the properties of the galaxy on larger scales. 
One may also argue that galactic escape fraction cannot be properly calculated without self-consistently considering the interaction of the ionizing radiation at molecular cloud scales.

Sometimes several SFMC particles cluster in or around a single computational cell and show nearly identical escape fractions.  
This phenomenon is manifested in Figure \ref{fig:age_vs_fesc}, where 18 particles have  $f_{\rm esc}(i) \sim 80\%$. 
This type of cell has ended up hosting multiple SFMC particles because it sits right at the center of a gravitationally unstable gas clump that potentially exhibits runaway collapse.  
This cluster of SFMC particles can be regarded as a large stellar nursery or a massive young star cluster of $\gtrsim 10^5 M_{\odot}$ that is under-resolved with our numerical resolution, 3.8 pc.  

Readers should also note that the escape fraction slowly rises as a function of particle age.  
By averaging 20 snapshots between 13.3 Myr and 30.7 Myr into the high-resolution evolution, we discover that $f_{\rm esc}(i)$ increases from 0.27\% at its birth to 2.1\% at the end of a SFMC particle lifetime, 6 Myrs.  
In other words, our best fit is
\begin{eqnarray}
f_{\rm esc}(i) = 0.0031\times T(i) + 0.0027
\end{eqnarray}
where $T(i) = t-t_{\rm cr}(i)$ is the age of the particle in Myr with particle creation time $t_{\rm cr}(i)$, as defined in \S\ref{sec-III:2-TF}.  
This result indicates that young SFMC particles are usually still buried in cold neutral gas in which they were spawned.  
As the SFMC particles age, they may drift away from the dense gas clump causing $f_{\rm esc}(i)$ to rise.  
The trend is also aided by ionizing stellar radiation and supernova explosion feedback (peaking at 5 Myr marked with a vertical line in Figure \ref{fig:age_vs_fesc}) which disperse the gas around the particle.  

\begin{figure*}[t]
\epsscale{0.79}
\plotone{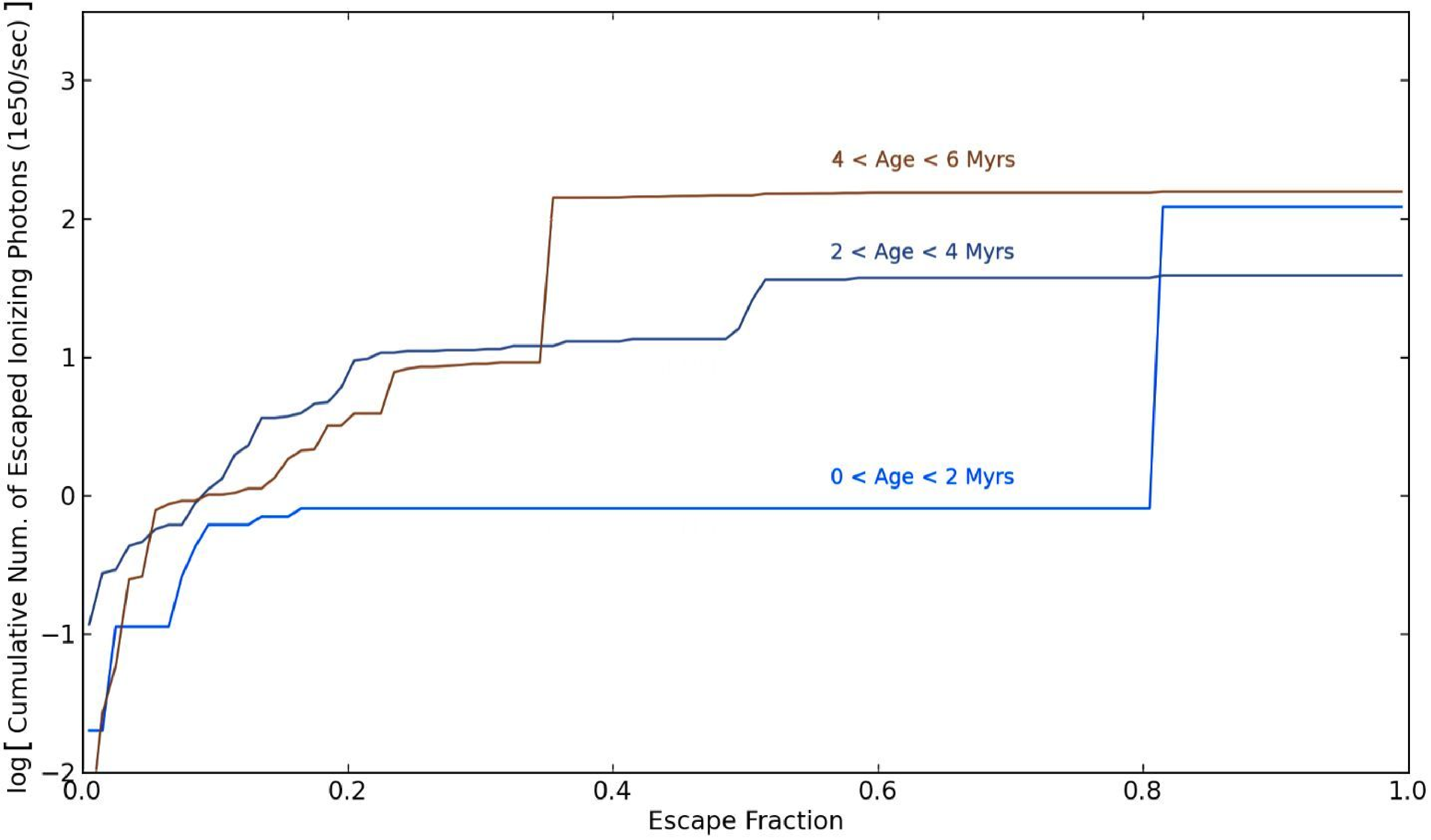}
    \caption{The cumulative numbers of escaped ionizing photons as a function of escape fraction per SFMC particle, $f_{\rm esc}(i)$.  Derived at 29.9 Myr into the high-resolution evolution. Three different age bins are shown.  For example, for the age bin of [0, 2] Myr, particles with $f_{\rm esc}(i) \sim 80\%$ contribute the most to the escaped photons in this bin.  In all three age bins, contributions by particles with $f_{\rm esc}(i) < 20\%$ are minimal.  More examination of this plot is provided in the text. 
\label{fig:fesc_vs_photons}}
\end{figure*}

Because older SFMC particles tend to have higher $f_{\rm esc}(i)$, they contribute equally or more to the total escaped photons despite the fact that they radiate fewer photons at the outset.  
Note that $M_{\rm MC}(i, t)$ in Eq.(\ref{eq:lum_MC}) is a function of time decreasing sharply from $M_{\rm MC}^{\rm init}$ at $T = 4$ Myr to $0.2 \,\,  M_{\rm MC}^{\rm init}$ at $T = 6$ Myr.  
The radiation luminosity, $L_{\rm MC}(i,t)$, experiences the same 80\% reduction in the last 2 Myrs of SFMC particle lifetime. 
However the increase in $f_{\rm esc}(i)$ typically offsets the decrease in $L_{\rm MC}(i,t)$, so the number of escaped photons scarcely depends on the particle age.
In order to illustrate this discovery, cumulative numbers of escaped ionizing photons are drawn as a function of $f_{\rm esc}(i)$  in Figure \ref{fig:fesc_vs_photons}. 
For all SFMC particles at 29.9 Myr into the high-resolution evolution, we compute both the number of ionizing photons which escape from that particle, and $f_{\rm esc}(i)$. 
We then break the particles into three age bins of [0, 2] Myr, [2, 4] Myr, and [4, 6] Myr. 
Within each bin, we order the particles by escape fraction, from lowest to highest, then compute how many escaping photons come from particles with that value of $f_{\rm esc}(i)$ or less, and plot the result in Figure \ref{fig:fesc_vs_photons}.
At this particular timestep, the contribution to the total escaped photons by the particles of age [4, 6] Myr is the greatest, close to a half of the total number. 
It is followed by the other age bins, [0, 2] Myr and [2, 4] Myr. 
Again, this phenomenon is in spite of the fact that the particles of age [4, 6] Myr originally radiate the least number of photons. 
Our result bears a striking resemblance to the radiation hydrodynamics simulation result by \cite{2012MNRAS.424..377D}, where clusters with the lowest initial luminosities have the highest escape fractions, and vice versa.  
These observations suggest that the escaped stellar photons may at least provide a sufficient number of photons to sustain galactic DIG layers.

Lastly, various other useful information can also be extracted by carefully examining this figure.  
For example, for the age bin of [0, 2] Myr, particles with $f_{\rm esc}(i) \sim 80\%$ contributes the most to the escaped photons in this bin.  
In contrast, the contribution by particles with $f_{\rm esc}(i) < 75\%$ is less than 1\%. 
In all three age bins, the contribution by particles with $f_{\rm esc}(i) < 20\%$ is no more than $\sim$ 5\%.  
This assessment reaffirms the claim that the escaped photon flux from a galaxy is predominantly controlled by a small number of SFMC particles with exceptionally high $f_{\rm esc}(i)$.  
It also is consistent with the observational findings by \citet[][Fig. 21]{2012ApJ...755...40P}.  

\subsection{Comparison to the Run Without Stellar Radiation} \label{sec-III:5-comp}

\begin{figure*}[t]
\epsscale{1.2}
\plotone{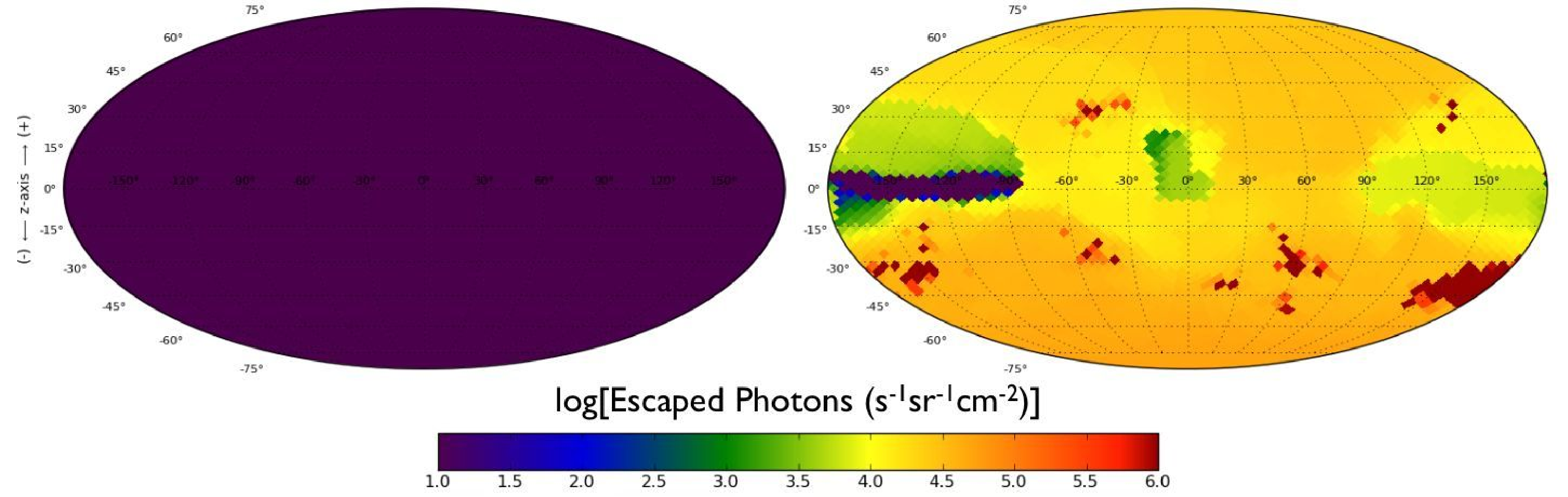}
    \caption{The escaped ionizing photon fluxes measured at 30.7 Myr into the high-resolution evolution for the MC-TF ({\it left}, only with supernova feedback) and MC-RTF runs ({\it right}, with both supernova and stellar radiation feedback).  The average escape fraction is $f_{\rm esc} = 0.00\%$ and 0.56\%, respectively.  See the caption of Figure \ref{fig:fesc_clump_allsky} for detailed information about this projection.  For the MC-TF run the map shows what the photon fluxes would have looked like had the SFMC particles begun to radiate at this very moment; note that for this run, fluxes in all pixels are considerably lower than the lower bound of the color bar.  Because SFMC particles are often buried deep in the cold neutral gas, photons would have rarely escaped the galaxy.     
\label{fig:fesc_compare_TF}}
\end{figure*}

We have thus far focused on the MC-RTF run, which includes stellar radiation feedback. 
However, it is very illuminating to compare the results to the MC-TF run, which includes supernova feedback but not radiation. 
To examine this, we have post-processed this run in exactly the same manner as the MC-RTF run. 
Surprisingly, the result of this exercise is that the escape fraction of the MC-TF run is negligible at all timesteps.
Figure \ref{fig:fesc_compare_TF} shows an example, comparing runs MC-TF and MC-RTF at 30.7 Myr into the high-resolution evolution. 
This difference is not because the behaviors of star formation in the two runs are radically different. 
Instead, it is because supernova feedback alone fails to provide sufficient ionization to create paths of low neutral fraction through which photons can escape. 
This is partly a matter of timing: by the time supernovae are going off and clearing the gas around SFMC particles, the ionizing luminosity is already in decline. 
It is also partly a matter of ionization budget: supernovae carry a lot of energy, but they tend to produce small amounts of hot gas but leave large quantities of neutral gas around them, and the neutral gas effectively blocks the escape of ionizing photons. 
The comparison of runs MC-RTF and MC-TF illustrates that one can only capture the escape fraction properly by self-consistently including photoionization as part of the calculation. 
Post-processing alone is insufficient.

\section{Summary and Conclusions} \label{sec-III:6}

We have performed a comprehensive high-resolution simulation of a dwarf-sized galactic disk including a sophisticated model of stellar feedback.  
Our goal has been to investigate galactic star formation and the evolution of a galactic disk in a single self-consistent framework, without omitting the detailed interaction of star-forming gas clumps with their surrounding galactic envelope.  
Our major findings are as follows.  

\begin{enumerate}
\item {\it Self-consistent Numerical Study of Galactic Star Formation:}
We have implemented a new way of describing feedback from SFMC particles in galactic simulations by combining ionizing stellar radiation and supernova explosion.  
The ionizing radiation feedback from each of $\lesssim$10000 SFMC particles is rendered by tracing the ultraviolet photon rays on the fly (\S\ref{sec-III:2}).  
Joined with high numerical resolution of 3.8 pc, the realistic description of stellar feedback helps to self-regulate star formation (\S\ref{sec-III:4-SF}).
Our simulation incorporates galactic gas envelope and SFMC particles in one numerical framework, and permits us to explore how galaxies and star clusters co-evolve and influence each other. 

\item {\it Escape Fraction via All-Sky Projection:}
By integrating the neutral hydrogen number density along different lines of sight we obtain the escape fraction of ionizing radiation from an individual SFMC particle or from a galaxy (\S\ref{sec-III:5-allsky}).  
Our post-production procedure does not include any additional radiative transfer, and is interactive and highly user-adaptive.
The escape fraction study in this work is possible only because three key components are consistently integrated with one another: {\it (a)} the realistic stellar feedback of ionizing radiation, {\it (b)} high spatial resolution and primordial chemistry to reproduce the galactic ISM, and {\it (c)} the post-production procedure to estimate the flux of escaped photons from an individual SFMC particle, or from a galaxy.
The machinery established in this work brings a unique insight into the physics of photon escape fraction.  

\item {\it Angular and Temporal Variations of Average Escape Fraction:} 
By simulating a low-redshift analogue of a galactic disk in the halo of $2.3 \times 10^{11} M_{\odot}$, we find that the average escape fraction  from all radiating sources on the spiral arms (excluding the central 2.5 kpc), $f_{\rm esc}$, fluctuates between 0.08\% and 5.9\% during a $\sim$ 20 Myr period with a mean value of 1.1\% (\S\ref{sec-III:5-time}).  
The flux of escaped photons from these sources is not strongly beamed, but manifests a large opening angle of more than $60^{\circ}$ from the galactic pole.  
It is often the smallest along the galactic disk due to the large column density of neutral hydrogen (\S\ref{sec-III:5-angle}). 
We however caution that these pictures might have changed if the SFMC particles in the inner disk had radiated ionizing photons  (Appendix \ref{sec:appendix-A}).  

\item {\it Escape Fraction Per Star-forming Particle:}
We also inspect the escape fraction per SFMC particle, $f_{\rm esc}(i)$.
We discover that the average escape fraction $f_{\rm esc}$ is dominated by a small number of SFMC particles with exceptionally high $f_{\rm esc}(i)$.  
One may therefore argue that the escape of ionizing photons is primarily determined by the small-scale properties of the clouds in the vicinity of a young star cluster, not by an overall galactic structure. 
On average, the escape fraction from a SFMC particle rises from 0.27\% at its birth to 2.1\% at the end of a particle lifetime, 6 Myrs.  
This is because SFMC particles drift away from the dense gas clumps in which they were born, and because the gas around the SFMC particles is dispersed by ionizing radiation and supernova feedback. 
Since older SFMC particles tend to have higher $f_{\rm esc}(i)$, they contribute equally or more to the total escaped photons despite the fact that they initially radiate fewer photons (\S\ref{sec-III:5-space}).  
\end{enumerate} 

The new feedback scheme described in this study opens numerous doors not only to understand the escape of ionizing photons from star-forming clumps, but to make mock observations such as H$\alpha$ emission and to examine the evolving environment of star-forming gas clumps.  
Using such mock observations, the companion paper in this series discusses the simulated star formation relations in dwarf-sized galactic disks, and how well they agree with observations \citep{2012arXiv1210.6988K}.   

\vspace{1 mm}

\acknowledgments

J. K. thanks Jeff Oishi, Joel Primack, and Chao-chin Yang for providing insightful comments and valuable advice.    
M. R. K. acknowledges support from an Alfred P. Sloan Fellowship, from the NSF through grant CAREER-0955300, and from NASA through a Chandra Space Telescope Grant and through Astrophysics Theory and Fundamental Physics Grant NNX09AK31G.
J. H. W. gratefully acknowledges support from the NSF Grant AST-1211626.  
M. J. T. gratefully acknowledges support from the NSF Grant OCI-1048505. 
N. J. G. is supported by a Graduate Research Fellowship from the NSF.
The examination of the simulation data and the post-production analysis are greatly aided by an AMR analysis toolkit {\it yt} \citep{2011ApJS..192....9T}.
This work used the Extreme Science and Engineering Discovery Environment (XSEDE), which is supported by NSF grant OCI-1053575.
The authors acknowledge the Texas Advanced Computing Center (TACC) at the University of Texas at Austin for providing high-performance computing resources that have contributed to the research results reported within this paper.
The authors are also grateful for the support from Stuart Marshall, Ken Zhou, and the computational team at SLAC National Accelerator Laboratory during their usage of the Orange cluster.

\begin{appendix}

\begin{figure*}[t]
\epsscale{1.17}
\plotone{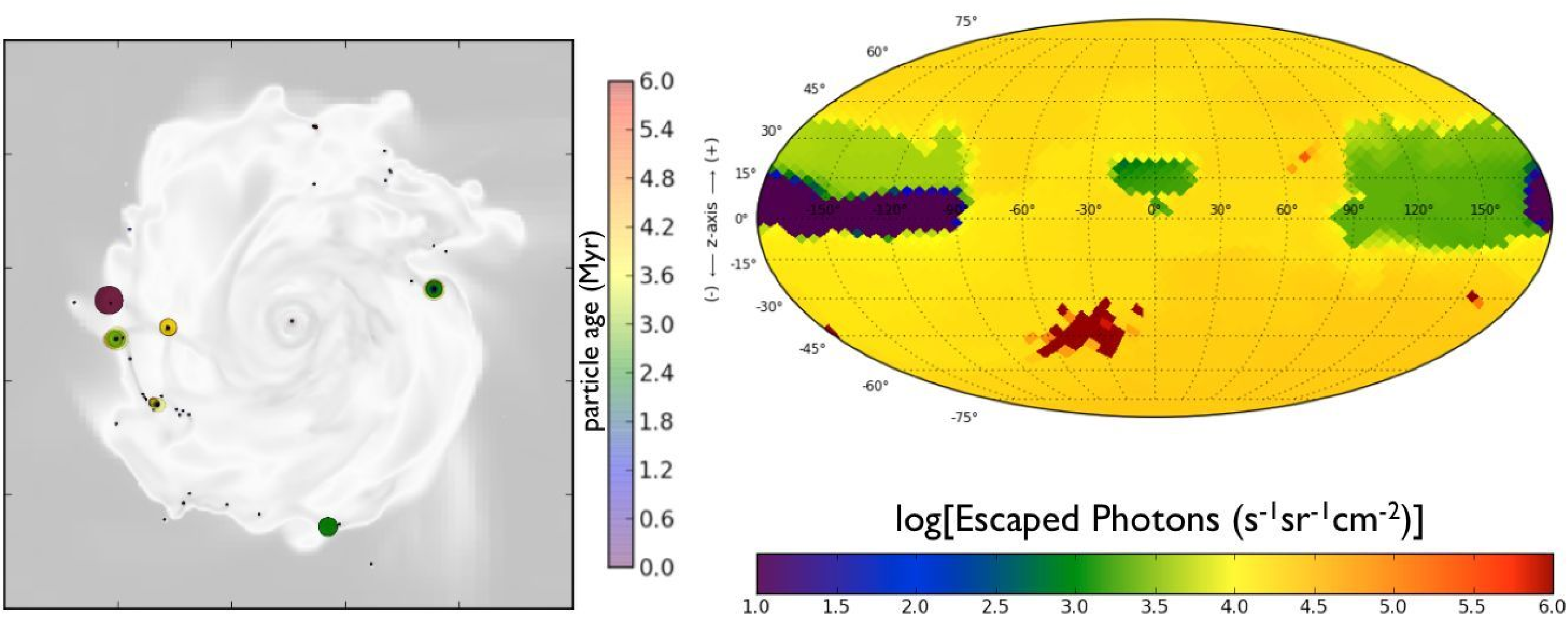}
    \caption{Same as Figure \ref{fig:fesc_clump_allsky} but for a simulation {\it without excluding} any SFMC particles from the radiation calculation.  {\it Left:} the locations of SFMC particles sized by their escape fractions, $f_{\rm esc}(i)$, and colored by their ages.  $f_{\rm esc}(i)$ values are evaluated at 28.1 Myr into the high-resolution evolution.  Note the SFMC particles in the inner galactic disk which now participate in the radiation calculation.
    {\it Right:} the weighted sum of the fluxes of escaped photon numbers from SFMC particles along different lines of sight, and seen at 100 kpc from the ray-emitting sources.  The average escape fraction at this snapshot is $f_{\rm esc} = 0.37 \%$.  See the caption of Figure \ref{fig:fesc_clump_allsky} for a detailed description of the coordinate system. 
\label{fig:fesc_clump_allsky_inner}}
\end{figure*}

\section{A. Effect of Including The Inner Disk on Escape Fraction Analysis} \label{sec:appendix-A}

In this section we investigate the effect of including the inner disk on the estimate of the averaged escape fraction, $f_{\rm esc}$.
As mentioned in \S\ref{sec-III:2-RF} and \S\ref{sec-III:5-time}, in the presented experiment only the SFMC particles that are more than 2.5 kpc away from the galactic center radiate ionizing photons. 
This treatment aims to exclude the densest portion of stellar disk which could be the result of an incorrectly structured central mass concentration.
This way, we can also concentrate on the evolution of the SFMC particles in the galactic spiral arms and outer disk.   
An added benefit is that the radiative transport calculation is expedited, although the improvement in computational cost is only $\sim10\%$ during the course of the calculation.

To investigate whether our choice of exclusion would introduce any significant bias in the reported escape fraction analysis, we have performed another simulation similar to the MC-RTF run, but without excluding any SFMC particles from the radiation calculation.  
We measure the average escape fraction from all radiating sources on the galactic disk, $f_{\rm esc}$, at four snapshots at 17.0 ($f_{\rm esc}$ = 0.53\%), 20.7 (2.1\%), 24.4 (1.4\%), 28.1 (0.37\%) Myr into the high-resolution evolution, equally spaced by 3.7 Myrs. 
We discover that the time-averaged mean of $f_{\rm esc}$ is 1.1\%, different by less than a percent from the estimate of the MC-RTF run (see \S\ref{sec-III:5-time}). 

In addition, Figure \ref{fig:fesc_clump_allsky_inner} visualizes the locations of SFMC particles with high $f_{\rm esc}(i)$ values, and the flux of escaped photons in the same manner as Figure \ref{fig:fesc_clump_allsky} but for the simulation without excluding any SFMC particles. 
On the left, one may note the SFMC particles in the inner galactic disk which now participate in the radiation calculation, but whose sizes are small in proportion to their $f_{\rm esc}(i)$ values. 
We find that these particles do not introduce any drastic change in the distribution of SFMC particles with high $f_{\rm esc}(i)$ values, nor in the beaming of the escaped photon flux.  
On average they do not have particularly low or high $f_{\rm esc}(i)$ values, the quantity which we are most interested in.  
In other words, there is nothing special about the SFMC particles in the inner disk.  

In conclusion, as long as we focus on understanding the {\it escape fraction} of ionizing photons from star-forming clumps and from a galactic disk, our choice of excluding the SFMC particles that are less than 2.5 kpc from the galactic center can be justified.  
It is even desirable to exclude such regions of a simulated disk that could have resulted from an incorrect central mass concentration  (see \S\ref{sec-III:2-RF}).
Evidently, however, this choice inhibits us from evaluating the {\it total number} of escaped photons from a galaxy. 

\section{B. Effect of Including Dust on Escape Fraction Analysis} \label{sec:appendix-B}

In this section we investigate the effect of including dust extinction on the estimate of the escape fractions of SFMC particles, $f_{\rm esc} (i)$.
While \cite{2008ApJ...672..765G} already demonstrated that the escape of ionizing photons above the Lyman limit is not heavily affected by dust, this experiment is to reexamine the case for ionizing photons we employed, $E_{\rm ph} = 16.0\, {\rm eV}$.
To this end, we first modify the escape fraction along different lines of sight, Eq.(\ref{eq:directional_escape_fraction}), in our post-production procedure
\begin{eqnarray}
f_{\rm esc}(i, \theta, \varphi) = e^{-\tau(i, \theta, \varphi)}
\end{eqnarray}
by adding the dust optical depth $\tau_{\rm d}$ as
\begin{eqnarray}
\tau = \tau_{\rm H, 100\,\, kpc} + \tau_{\rm d, 100\,\, kpc} = \sigma_{\rm H} \Sigma_{n_{\rm H}, 100\,\, {\rm kpc}}  +\sigma_{\rm d} (\Sigma_{n_{\rm H}, 100\,\, {\rm kpc}}  + \Sigma_{n_{\rm H^+}, 100\,\, {\rm kpc}}).
\end{eqnarray}
Here $\sigma_{\rm d} = 10^{-21}\,Z\,'$ is the mean extinction cross section by dust per hydrogen nucleus with the metallicity $Z\,'$ normalized to the solar value \citep{2011ApJ...729...36K, 2012ApJ...749...36K}.  
We use the local metallicity in the cell through which the optical depth is computed.
Readers should note that even though $\sigma_{\rm d}$ is scaled with the metallicity of the gas, our choice of $\sigma_{\rm d}$ fits well only in the Milky Way-like galactic ISM.  
Because $\sigma_{\rm d}$ does not depend on temperature, nor does it consider the short lifetime of dust grains in a hot gaseous halo, it would very likely {\it overestimate} the optical depth through the halo, especially along the lines of sight that are perpendicular to the disk plane.  
That is to say, our choice of $\sigma_{\rm d}$ is very conservative for our purpose.  

\begin{figure*}[t]
\epsscale{1.16}
\plotone{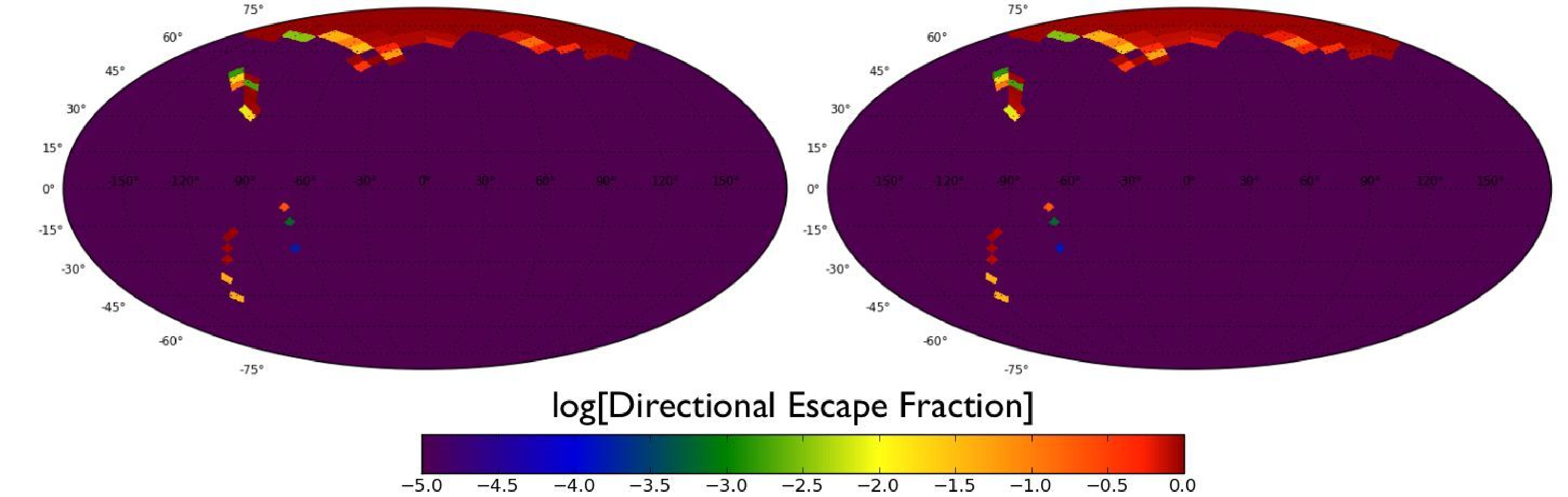}
    \caption{The all-sky map of the photon escape fraction at 100 kpc along different lines of sight from a SFMC particle, $f_{\rm esc}(i, \theta, \varphi)$, without ({\it left}) and with ({\it right}) the effect of dust.  They are measured at 29.9 Myr into the high-resolution evolution in the MC-RTF run.  The sky-averaged escape fraction from this particle is $f_{\rm esc}(i) = 5.5\%$ and 5.2\%, respectively.  The difference between the two images introduced by dust is barely noticeable in the polar region.
   \label{fig:fesc_compare_dust}}
\end{figure*}

With this revised formula we reevaluate the photon escape fraction in the MC-RTF run.  
From this experiment we find that dust extinction reduces the escape fraction only by a few percent even with our conservative choice of $\sigma_{\rm d}$, in clear agreement with \cite{2008ApJ...672..765G}.  
For example, Figure \ref{fig:fesc_compare_dust} shows the all-sky map of the escape fraction along different lines of sight from a SFMC particle, $f_{\rm esc}(i, \theta, \varphi)$, without and with the effect of dust.  
While some careful readers might notice the slight difference between the two images, the overall sky-averaged escape fraction is reduced only by $\sim5\%$, from $f_{\rm esc} (i) = 5.5\%$ to 5.2\% by the inclusion of dust.  
Considering that this difference is still an overestimation due to the lack of more accurate parameterization for $\sigma_{\rm d}$, we conclude that dust extinction does not significantly revise the escape fraction estimates for ionizing photons.
We therefore choose not to include the effect of dust in the radiation hydrodynamics and the post-process (see \S\ref{sec-III:2-RF} and \S\ref{sec-III:5-allsky}).   

\end{appendix}

\end{document}